\documentclass{hcij}
\bibliography{./hcij_article.bib}

\usepackage{graphicx}
\usepackage{multirow}
\usepackage{float}
\usepackage{etoolbox}
\usepackage[normalem]{ulem}

\usepackage[utf8]{inputenc}

\begin{document}

\articletitle{Bridging Social Distance During Social Distancing: Exploring Remote Collegiality in Video Conferencing}

\runninghead{Bridging Social Distance During Social Distancing}

\articleabstract{Video conferencing systems have long facilitated work-related conversations among remote teams. However, social distancing due to the COVID-19 pandemic has forced colleagues to use video conferencing platforms to additionally fulfil social needs. Social talk, or informal talk, is an important workplace practice that is used to build and maintain bonds in everyday interactions among colleagues. Currently, there is a limited understanding of how video conferencing facilitates multiparty social interactions among colleagues. In our paper, we examine social talk practices during the COVID-19 pandemic among remote colleagues through semi-structured interviews. We uncovered three key themes in our interviews, discussing 1) the changing purposes and opportunities afforded by using video conferencing for social talk with colleagues, 2) how the nature of existing relationships and status of colleagues influences social conversations and 3) the challenges and changing conversational norms around politeness and etiquette when using video conferencing to hold social conversations. We discuss these results in relation to the impact that video conferencing tools have on remote social talk between colleagues and outline design and best practice considerations for multiparty videoconferencing social talk in the workplace.}

\newpage

\thickhrrule

\tableofcontents

\thickhrrule

\articlebodystart

\section{Introduction}
Casual conversation, where people engage in social talk or phatic communion \parencite{coupland1992how} (i.e., non-task oriented talk) is acknowledged as important in facilitating collaboration among colleagues \parencite{kraut1990informal}. Workplace interactions of this nature foster stronger social bonds \parencite{coupland2014small} and contribute to increased job satisfaction \parencite{riordan1995opportunity}. Before the COVID-19 pandemic, these interactions would largely be driven by the physical environment, where proximity is a core indicator of whether these conversations occur \parencite{rockmann2015contagious,kraut1990informal}. Supporting this, research has noted a lack of opportunity for casual interactions within virtual teams \parencite{rockmann2015contagious}, with increased feelings of social distance \parencite{robert2018disaggregating} and loneliness \parencite{macik2006virtual}. The recent move to working from home (WFH) due to COVID-19 restrictions means that social talk between groups of colleagues has moved online - using video conferencing (VC) tools like Zoom and Microsoft Teams - transforming how colleagues \emph{do collegiality}. As WFH becomes more common post-pandemic, important questions arise about how computer-mediated communication (CMC) tools such as VC applications can best be used to facilitate social talk between groups of colleagues. 

In this paper, we report on a study conducted during August and September 2020, which aimed to examine how people have maintained social interaction and engaged in social talk with their colleagues while working from home. Participants were asked to engage in a multiparty (3-way) social conversational session with colleagues. After this, each participant completed a semi-structured interview reflecting on their remote social interactions with colleagues while working from home, as well as their experience in the study. 

We find that VC influences the dynamics and topics of workplace social conversations, and the role these conversations play in maintaining pre-existing relationships and forming new ones. Through our data analysis we identify three core themes that emphasize 1) the purposes and opportunities that VC allows when conducting social talk with colleagues, 2) the role that professional and personal relationships play in social conversations, and 3) the politeness and etiquette norms that have formed through using VC to hold social conversations with multiple colleagues. From this, we outline design considerations and best practices for social talk applications in VC that explicate how these platforms can best be used to support remote collegiality.

\section{Related Work}

\subsection{Social Talk \& Workplace Communication}
Social talk in the workplace, often glossed as 'watercooler conversation', is associated with increased workplace well-being \parencite{methot2020office}. Although these interactions are sometimes seen as a source of interruption \parencite{jett2003work}, social talk between colleagues in the workplace is considered a prosocial behavior \parencite{george1990understanding} that provides a number of key benefits. It is critical to the development of social bonds and interpersonal relationships between colleagues \parencite{coupland2014small}. It also has a positive impact on work-related outcomes, such as meeting success \parencite{allen2014linking}, perceived workplace opportunities \parencite{lin2015potential} and decision making in organizations \parencite{sproull1984nature}. Social talk topics range from the weather, to gossip, humor, and other aspects of everyday life \parencite{coupland1992how}, but also includes exchanges of organizational information and social support \parencite{brass1984being,ibarra1992homophily}. It also allows colleagues to transition between difficult and serious subjects \parencite{knutson1986exploration}, commonly occurring before, or during the opening and closing phases of meetings \parencite{mirivel2005premeeting}. Organizations often make an active effort to foster social talk between colleagues, such as through the scheduling of company events \parencite{tews2013does} and the creation of shared common spaces \parencite{nejati2016implications} where spontaneous and opportune moments for social talk are able to occur \parencite{whittaker1994informal}. 

All talk relies on turn taking, adjacency, and participation shifts \parencite{sacks1978simplest,gibson2003participation}. Social talk is made up of phases such as \emph{chat segments} where there are multiple switches in turns, and \emph{chunks} wherein one participant dominates the floor for an extended period of time (e.g. storytelling) \parencite{gilmartin2018explorations}. During these social conversations, interlocutors tend to take on dynamic roles, switching speaker roles more frequently than in task-based dialogues, as partners collaborate to maintain social bonds while avoiding awkward silences \parencite{gilmartin2017exploring}.
Although the line between social talk and professional or `institutional' talk can seem blurred, there are clear distinctions in form, which previous work has studied and which we briefly describe here to clarify the important differences.
Institutional talk is often constrained by pre-determined systems of turn-taking based on speaker status, or organizational protocols. For example, \cite{heritage2010talk} categorize `turn-type pre-allocation systems' such as interviews or classrooms with a clear hierarchy, and `mediated turn-taking systems' such as committee meetings where a moderator orchestrates the turns. In contrast to social talk, turn-taking behaviors are rarely present unless the context has been specifically tailored for their acceptability \parencite{gibson2003participation}, or protocol is being actively challenged \parencite{ford2008women}.

Much of the research on social talk in the workplace focuses exclusively on in-person rather than remote work contexts, with the exception of work on social communication between distributed physical teams (e.g., \parencite{fish1990videowindow}). Work on social talk in VC is currently more common around domestic settings. In the home, VC applications are adopted to help bridge the distance between couples, family, and friends \parencite{brubaker2012focusing,rintel2013video,judge2010sharing}, allowing them to stay in touch \parencite{forghani2014routines,judge2010sharing} check in on one another, and share joint activities. Due to the COVID-19 pandemic, many organizations are now looking to use VC tools to support and socially engage employees and work teams. To the best of our knowledge, how VC technologies are used to maintain collegial relations between team members and employees has yet to be examined in detail. 

\subsection{Video Conferencing and its influence on Conversational Processes}
It is clear that conversational behavior differs in VC when compared to in-person communication. In particular, studies \parencite{o1993conversations,sellen1992speech,sellen1995remote} have investigated how conversational structure and linguistic phenomena change during VC. Such work has focused on simultaneous speech \parencite{sellen1992speech, sellen1995remote}, backchanneling \parencite{o1993conversations}, and the length, distribution, and number of turns \parencite{sellen1992speech,sellen1995remote} in work-related conversations that focus on conducting a defined task. In these contexts, compared to in-person interactions, VC tends to lead speakers to interrupt less, while also leading to a reduction in backchanneling \parencite{sellen1995remote,o1993conversations}. These findings are thought to demonstrate an increased formality in VC settings. Interlocutors also experiences difficulties in turn taking when using VC \parencite{sellen1995remote,o1993conversations}, which may be in part explained by perceived difficulties in clearly establishing common ground. This issue may be particularly acute in virtual multiparty conversations. In studies manipulating group size in VC settings, acquiring a basis for common ground has been shown to require more conversation effort through increased conversational turns as groups increase in size \parencite{anderson2006achieving,anderson1999understanding}. Although there is some work on multiparty conversations on VC tools \parencite{o1993conversations,sellen1995remote}, this work is once again around task-based interactions, with little focus on how multiparty social talk is conducted and influenced by VC.

\subsection{Technologies Supporting Social Talk in the Workplace}

Past work has defined in-person social talk in the workplace as dyadic, brief, and unplanned, with spontaneous conversational participants \parencite{whittaker1994informal,kraut1990informal}. Studies in the CMC domain have sought to understand how these interactions are supported in-person through instant messaging, suggesting that such tools play a role in initiating social talk in the workplace, termed \emph{outeraction} \parencite{nardi2000interaction}. Indeed, awareness of conversational participants' availability is key to facilitating social talk initiations. However, existing collegial bonds, oftentimes solely a consequence of in-person workplaces, are a prerequisite to negotiating availability for these conversations \parencite{nardi2000interaction,nardi2002place}. Moreover, \emph{communication zones} must be available to allow for conversational participants to intermittently engage in informal communication and maintain communication context in these brief interactions, comparable to the always-on video links present in media spaces \parencite{nardi2000interaction}. In these zones, communication takes place between colleagues who have existing bonds, formed by previous communicative activities \parencite{nardi2002place}. However, despite the support of both multiparty and dyadic interactions in social talk applications in the workplace \parencite{nardi2000interaction}, dyadic social talk remains the default form of informal conversation studied \parencite{whittaker1994informal,nardi2000interaction,whittaker1995rethinking}.

More recently, literature describing the effects of the pandemic points to the scarcity of social talk in the workplace \parencite{kaushik2020impact}, outlining issues such as isolation, stress, and the lack of clear work-life boundaries \parencite{de2020recover}. The rapid shift to fully remote work has outlined issues with fatigue and the need to take breaks in longer conversations in VC \parencite{cao2021large} as well as the lack of well-balanced discussions due to limited non-verbal cues \parencite{cao2021my}. What is more, the shift to remote work has further complicated existing work-life boundaries, as workers plan their days more carefully \parencite{koehne2012remote} and are more task-focused when working remotely \parencite{yang2020work}. This may be an effort from workers to prove to themselves and their managers that working from home is productive \parencite{halford2005hybrid}, resulting in conversations that are directed at work-related rather than social goals. Coupled with the longer working hours reported during the pandemic \parencite{rudnicka2020eworklife}, many employees report challenges in creating clear divisions between their work and personal lives, creating risks such as burn-out \parencite{hayes2020m}. Hence, opportunities to engage in social conversations may prove essential to maintain qualities of an in-person workplace, providing a sense of collegiality during remote work to reduce the aforementioned risks \parencite{doolittle2021association}.

Although the COVID-19 pandemic has increased the visibility of WFH practices, with a recent wave of associated publications, the focus is on productivity \parencite{cao2021large,sarkar2021promise}, and essentially examines remote meetings rather than socializing remotely. 
Past work \parencite{stray2019slack} has shown that team tools such as Slack may provide opportunities for social talk alongside task-based information exchange in remote teams. In addition to teams leveraging features such as parallel chat in VC to engage in social conversations \parencite{sarkar2021promise}, 
communication however often remains problem-focused in both remote and in-person workplace environments \parencite{calefato2020case,stray2019slack}. With increased communication on tools such as Microsoft Teams during the pandemic \parencite{microsoft:pandemic}, understanding ecosystems that support social conversations become evermore important, as the frequency of communication has increased, and will likely continue, as remote work becomes more embedded in work practices. However, a clear understanding of how social talk occurs in remote workplace environments and how technologies support these interactions has, to the best of our knowledge, yet to be reached.

\section{Current Study}
The COVID-19 pandemic has led to a significant growth and interest in WFH practices. Many companies are now having to use VC tools to facilitate social interactions between employees that would have occurred in person prior to the pandemic. With increased WFH likely to define the future of work, understanding how social talk is being conducted through VC is important to identify issues and opportunities on how to best support this activity. In our work, we evaluate the practice of social talk and the role it takes on in remote work settings, and consider WFH factors that influence how colleagues engage in social talk, in addition to  
how VC technologies support multiparty social talk conversations. Building on recent work during the COVID-19 pandemic, past work on workplace studies of dyadic in-person social talk \parencite{nardi2002place,whittaker1994informal,whittaker1997telenotes}, task-based multiparty CMC \parencite{cohen1982speaker,sellen1992speech,sellen1995remote}, and communication in domestic and professional settings \parencite{brubaker2012focusing} we aim to understand people's perceptions of the nature and role of casual multiparty social talk between colleagues over VC as they work from home. For instance, what technologies and activities are used to facilitate VC multiparty social talk, and more importantly, what factors influence how colleagues choose to engage on these platforms? The contribution of our work lies in defining how colleagues stayed connected while working remotely. Although past work has examined social talk in mediated technologies in the workplace \parencite{nardi2000interaction}, to the best of our knowledge, an understanding of how VC facilitates these interactions among distributed colleagues has yet to be examined. 

\section{Method}

\subsection{Participants}
Ten groups of English speakers were recruited from five research institutes in three different countries (The Netherlands, Austria, Ireland), via email and snowball sampling. Recruitment and data gathering occurred during August and September 2020. Each group contained three members (N=30, W=16, M=14; Mean age=32.13 yrs; SD= 10.3 yrs). Based on previous literature \parencite{tang2014techniques} on recruiting participants for VC studies, we first recruited one participant by the aforementioned methods and then asked them to recruit two other colleagues in their organization. All participants were either native or near-native English speakers and/or used English as a common language at work. This was moderated with the call for participants and the demographic questionnaire in our study. To control for supervisor and subordinate relations, only participants at a similar career stage and rank took part in the study together. Additionally, to ensure that participants could reflect on the comparison between in-person and VC-based social talk during the interviews, all participants had worked with their group members in the same physical work environment prior to COVID-19. 

Of the 30 participants, 30\% (N=9) reported using video conferencing platforms daily for socializing purposes, with 43\% (N=13) using them multiple times a week, and 27\% (N=8) using them once a week or multiple times a month. In the sample, participants reported using multiple VC tools, with 97\% (N=29) of participants having used Zoom, 67\% (N=20) reported using Skype, 40\% (N=12) Microsoft Teams, and 27\% (N=8) using Google Hangouts, among other tools. 

\subsection{Conversation Session} To support reflection and stimulate discussion during the semi-structured interview sessions, each triad was asked to engage in a 20 minute free-form social conversation over Zoom. We acknowledge the lower ecological validity of our study due to the allocated time frame and recording of the conversation. However, we provide the following rationale for our decision: 1) 20 minutes was chosen to allow colleagues to shift through various topics, and 2) prior literature \parencite{rintel2007maximizing} has shown that people soon become comfortable with recordings on video conferencing after such time. To simulate social talk, participants were given no guidance as to what to discuss. Before the session, they were told that they were to converse on Zoom, unobserved, for approximately 20 minutes before being interrupted by the experimenter and provided with links to be interviewed in separate Zoom rooms. Participants were given instructions to converse "as they they would when meeting colleagues at the coffee machine or water-cooler for a casual chat". See figure 1 for an example of the setup (pg.38).

\subsection{Qualitative Data}
\subsubsectionnonum {Semi-structured Interviews}  After the conversation session, participants were asked questions that prompted them to reflect on: 1) their experience of in-person social talk relative to social talk on VC tools; 2) perceived conversational roles and dynamics during the session; 3) previous uses of collaborative technologies for social talk; and 4) common topics of discussion in social talk over VC. The interview was designed to ask participants about their recent social talk experience among colleagues in the last months, in addition to the experience of conversing in the planned 20 minute session. We paid attention to challenges of remote social talk and how communication was maintained with colleagues when working from home. Participants were asked to draw from their recent experience using VC tools during lockdown in addition to the recent conversational session. 

\subsection{Procedure}
The study was approved by the University’s low risk project ethics procedures [HS-E-20-93]. The conversational sessions were scheduled through Doodle and a link to the Zoom session was provided in a Google Calendar invite. Prior to the interview, participants received a participant information sheet with details concerning the study. At the scheduled time, participants joined individually, a multiparty Zoom session set up for the study. After completing a brief demographic questionnaire, participants then engaged in an unstructured conversation for 20 minutes, which was recorded. A segment of 20 minutes was chosen, following prior work \parencite{o1993conversations} using the same time frame. Directly after the conversation session, each participant received separate Zoom links to simultaneously take part in a 30-minute semi-structured interview. Participants were then debriefed as to the aims of the research and thanked by receiving a \texteuro15 voucher as a honorarium. 

\section{Results}

\subsection{Qualitative Analysis Approach}
Transcribed semi-structured interview data was analysed using inductive thematic analysis \parencite{guest2011applied}. Initial coding was conducted independently by two researchers. An inductive approach was used to group initial themes through NVivo 12.6.0. Afterwards, the researchers collaborated via Miro\footnote{https://miro.com/} to cluster initial themes identified. After discussing the independently coded themes, areas of disagreement were resolved by either collapsing or further refining them. The sections in the qualitative analysis represent the themes discussed and agreed on by the two researchers, both of whom have a prior background in HCI and previous experience conducting thematic analyses. 

Through the analysis, three core themes were identified: 1) ``Purpose and Opportunities''; 2) ``Relationships'' and 3) ``Etiquette and Politeness''. Within these themes a further 11 sub-themes were also identified (see Figure 2, pg.39). 

\subsection{Purpose and Opportunities}
Participants mentioned increased opportunities for social talk during the pandemic, as barriers to social talk were lowered with video conferencing. Although spontaneity of social talk in physical spaces was missed, participants benefited from interactions with colleagues whom they did not previously share close physical proximity with - an apparent prerequisite to frequent social conversations. Experiences among colleagues varied depending on the manner in which social talk was organized. 
 
\subsubsectionnonum{Media Ecosystems in Remote Social Talk}
Participants reported increased use of video conferencing tools for conversation with colleagues since the pandemic. Indeed, many reported using them every day in collaborative processes. These tools were mainly used for professional purposes, with many mentioning that social talk normally occurred at the beginning or end of professional meetings; as an aside to the task at hand.  

\begin{quote} 
\textit{``I would say at the moment and certainly since the COVID situation, it would be several times a week to speak to colleagues using tools like Zoom.''}-[P23]
\end{quote}

However, frequency of use differed depending on the relationship between colleagues, with one participant noting: \emph{``at least once a day, at least, though not all colleagues. Colleagues that I have good relationships with''} [P29]. Although messaging platforms were used to organize social talk meetings - as touched on by \cite{nardi2000interaction} - participants tended to use these with colleagues that they felt closer to, and used video conferencing platforms for colleagues they only had working professional ties with. 

\begin{quote} 
\textit{``We do not use chats too much. These we use rather with the private. Let's say with the family and close friends, but with colleagues, we mostly use this teleconferencing platforms and emails.''}-[P27]
\end{quote} 
 
In tandem with participants' increased use of VC, asynchronous communication tools have been increasingly adopted, or repurposed, as facilitators of social talk. For example, although not commonly mentioned, emails were often used to initiate planned social meetings.

\begin{quote} 
\textit{``We have a rolling weekly email that kind of sets up our, our coffee, a sort of our lunchtime afternoon coffee talk once per week.''}-[P18]
\end{quote}

Further, some team tools (e.g., Slack) and messaging platforms were mentioned in our interviews as places where social talk occurred, as they \emph{''encouraged spontaneous communication''} [P28] and facilitated \emph{''banter''} [P29], with these platforms exclusively \emph{''reserved for social talk''} [P25]. However, VC platforms were the default means of communication among colleagues, oftentimes used alongside shared collaborative tools (e.g., Google Docs, Miro, etc.) in task-based interactions such as brainstorming and collaborative writing. This may explain the prominence of social talk interactions on these platforms as colleagues \emph{''always find ways to talk, socially''} [P7].

Likewise, the formality of a video meeting tends to influence the platform through which it is organized, with email being seen as a tool for organizing more formal meet-ups compared to messaging tools such as WhatsApp.

\begin{quote}
\textit{``usually for the formal [video call] with our supervisor be over email and then for the informal social ones with the other members of the group would usually be over WhatsApp''}-[P14] 
\end{quote}

\begin{quote}
\textit{``You have to plan that, so that would nearly always be planned by email. So it would be for more of the professional meetings, [..] somebody has an issue or problem that they want to discuss and they'd say - they'd ask on an email. Yeah. Whereas Microsoft Teams would be more ad hoc''
}-[P29]
\end{quote}

Thus, the increased use of video conferencing tools has been coupled with an increased use of tools that enable initiation of video calls, which are  largely influenced by their purpose and formality.

\subsubsectionnonum{Opportunity for Social Talk}
Lack of chance encounters while working remotely was mentioned by participants as a negative consequence of WFH. Social conversations now had to be organized to occur, with one participant summarizing that \emph{``we have to schedule fun now''} [P8]. This was contrasted with the experience from in-person encounters where one could simply \emph{``walk by somebody's open door and you see how busy they are. And if they're not so busy you pop in for a chat''} [P29]. Despite the new means and methods inherent in WFH, the adherence to schedules, agendas, and deadlines, representative of the traditional workplace, has not changed.  There is thus some irony in having to treat the one element of workplace spontaneity as yet another calendar item, which has not been lost on participants. With no parallel for this in the remote world, scheduled fun has become the new norm for participants:

\begin{quote} 
\textit{``...therefore, you go, you know, this is now half an hour or whatever, it is predetermined to be social talk and in that respect, it's of course different from the accidental meeting over the photocopier or coffee''}-[P12]
\end{quote} 

\begin{quote} 
\textit{``It requires a certain intention like you intended to call each other, socially, which is a lot, you know, you're missing the serendipity of the - it's not spontaneous..''}-[P9]
\end{quote} 

Instigating predetermined social talk sessions was often done by colleagues who were \emph{"bored"} [P14] or felt a need for \emph{"non-work"} [P14] activities. Initiating these activities involved planning the spaces where the interaction would occur, in addition to checking on the availability of other colleagues. While attempting to replicate the natural gatherings of colleagues in the same physical proximity, this required more conscious effort from all involved.

\begin{quote} 
\textit{``a certain percentage of obstacle in terms of setting up the call, especially when compared to an interaction at the workplace ...''}-[P30]
\end{quote} 

Scheduled social `encounters' thereby become part of the working day for colleagues, with time allocated for them in between work tasks.

\begin{quote}
\textit{``it was an opportunity to kind of make space in our schedules, just to catch up and to see each other.''}-[P23]
\end{quote}

Despite the purposeful, scheduled nature of these chat calls, getting back to work was still a common concern among colleagues. As outlined in previous literature \parencite{jett2003work}, non-work talk was perceived as disruptive during work-related tasks. Participants reported that they often felt the need to be productive when working from home and felt they had to \emph{"get back to work"} [P21] rather than socialize. 

\begin{quote}
\textit{``when you're working at home, you feel you should be kind of working, you know, effectively and efficiently and to schedule.''}-[P23]
\end{quote}

In addition to informal social talk video calls with peers, participants mentioned formal occasions that are scheduled by senior members of their organizations, such as coffee mornings and scheduled `catch-up' sessions. 

\begin{quote}
\textit{``We started having coffee mornings to have one event that we, at least one, when we all come together and talk about what happened and what's coming up.''}-[P12]
\end{quote}

However, these interactions were considered to be \emph{"work-oriented"} [P18], with some noting (P17, P18) that these experiences were similar to work meetings with senior members of staff. 

\begin{quote}
\textit{``We have a weekly chat, which is organized by the quote unquote adults in the lab, which is kind of informal, but it still is something which the grown ups put together .''}-[P17]
\end{quote}

Outside of professional meeting-based small talk, social interactions were primarily organized events, with designated members of senior staff responsible for their scheduling and moderation. Although there were instances where colleagues would initiate social conversation calls for their peers with no prior plan, these were often brief and constrained due to pressure to be productive in WFH environments. 

\subsubsectionnonum{Building Remote Social Connectedness}
As entire organizations have been compelled to work from home, the otherwise implicit day-to-day routines and rituals of their members have too been disrupted. In response, video conferencing tools have been further applied to emulate such activities, with participants expressing a need to \emph{"compensate"} [P25] for their absence. 

\begin{quote}
\textit{``you know we tend to have a pattern, maybe a certain time of the day, maybe late afternoon, we would have those kind of calls to kind of substitute what we would have been doing anyway in the workplace..''}-[P30]
\end{quote}

\begin{quote}
\textit{``before the lockdown we would just meet in the hallway on campus. [..] Now after the lockdown obviously this possibility wasn't there anymore,  [..] so we set up these regular meetings.''}-[P6]
\end{quote}

As well as simple chats to substitute for in-person encounters, organized games and quizzes conducted through video conferencing provided structure to the conversation, and a welcome distraction that participants appreciated.

\begin{quote}
\textit{``I've been involved in ones where we've been doing quizzes or doing online games with each other as part of teams and I've had the same with family members that have been cocooning over the different periods and different locations, because we've got two frontline workers. So again, from that side of things having everyone on is excellent conversation.''}-[P15]
\end{quote}

Regular lunch and tea breaks were also reported among participants. For example, many mentioned experiences of commensality in joint experiences and planning them into their day.

\begin{quote}
\textit{``Online exactly. Just eat lunch together or rather the 20 minutes after we ate and talk about general stuff. So we usually arranged it before.''}-[P20]
\end{quote}

\begin{quote}
\textit{``There was a Teams, a Microsoft Teams meeting that was always open and people could go in from 9 to 9:15 and there would be someone there having a cup of coffee so you can start your day together..''}-[P9]
\end{quote}

Participants mentioned that organized social meetings allowed them to connect with colleagues outside their direct working group, increasing \emph{accessibility} [P29] and \emph{exposure} [P30] to social talk opportunities with coworkers outside their working environment.

\begin{quote}
\textit{``Thanks to these platforms, I found myself interacting with people. I never interacted before in person, because they weren't in my work sphere...''}-[P28]
\end{quote}

As well as stimulating social interactions between colleagues from across the office building, virtual meetings also provided an opportunity for some to connect with collaborators from across the world [P13,P15]. In one research group, colleagues saw VC as a way to interact with distant colleagues \emph{"without moving them"} [P13].

\begin{quote} 
\textit{``It was like a kind of a common sense. We, just after the first group meeting, we're like, okay, we are so far away from each other. I mean, my group, it's like people in Germany, France, Italy, Ireland. So we are really far away. So we were like, oh, we can see each other. So we need to do something...''}-[P13]
\end{quote}

Bonds that were forged pre-pandemic were also deepened, with some participants mentioning the increase of social versus work-oriented conversational topics. 

\begin{quote} 
\textit{``For some people, I've got to know them better because of COVID-19''}-[P30]
\end{quote}

\begin{quote} 
\textit{``everything about...our childhoods to the weather to, you know, our personal lives''} -[P15]
\end{quote}

Overall, participants used VC to create opportunities for remote social talk, enriching pre-existing bonds. Organized social talk activities, created to emulate in-person interactions, added back the pre-pandemic rituals of their day. Participants also saw the use of video conferencing as an opportunity to socialize with colleagues from other departments and areas of the office building that they would normally see relatively infrequently. Similarly, opportunities to interact with colleagues in different cities and countries also increased. This was often due to the planned nature of some of these interactions, allowing multiple colleagues to interact in larger groups. Although previous in-person rituals (e.g., eating or grabbing a coffee together) determined the nature of remotely organized social talk activities, the tendency to meet in larger groups was in contrast with prior in-person settings where dyadic and spontaneous encounters define social talk occurrences. 

\subsubsectionnonum{Talking Shop}
Despite the closer bonds that were forged during the pandemic, many participants still mentioned work talk creeping in to their social interactions with colleagues. One participant explained how \emph{``the very act of talking to a laptop makes me think about my job''} [P11]. More generally, these interactions were often defined as a mix of social and work-related interactions: 

\begin{quote} 
\textit{``Between deeply personal and deeply professional, then this would be somewhere in the middle.''}-[P16]
\end{quote}

\begin{quote} 
\textit{``There's no obligation to talk about anything to do with work, though inevitably something to do with work always comes up..''}-[P29]
\end{quote}

This was mainly because colleagues shared the same context, using these conversations for \emph{``releasing stress as well and guiding each other about challenges and, yeah, sympathizing with each other''} [P30], in addition to discussing COVID-related workplace changes. 

\begin{quote} 
\textit{``We were all basically just kind of sharing our experiences and, you know, gosh, is this going to work on campus.''}-[P24]
\end{quote}

\begin{quote} 
\textit{``What's going on at the workplace and in terms of people returning and different considerations around that and return to teaching this semester.''}-[P30]
\end{quote}

During the pandemic, mental health and checking in with colleagues also became an important topic of conversation, with many participants mentioning isolation as a key concern among colleagues.

\begin{quote} 
\textit{``There's a lot of checking in. Health obviously takes some precedence when doing small talk. Yeah, there's a lot of precarity and a lot of changes, also emotional and mental health have been making more of a comeback.''}-[P7]
\end{quote}

However, social talk topics often reverted back to COVID-related themes such as \emph{"the numbers of COVID are a bit high, or are not high"} [P2] with two participants reporting that at times conversations with some colleagues were exclusively about COVID due to fears: 

\begin{quote} 
\textit{``when you spoke to them it was all about COVID and you weren’t able to talk to them about anything else''}-[P20].
\end{quote}

\begin{quote} 
\textit{``I think nearly every social talk before meetings turns into a conversation about COVID on lockdown.''}-[P5]
\end{quote}

Overall, colleagues participated in similar chats to those in in-person environments, engaging in topics such as  \emph{``weather, sports, how everybody is getting on mentally''} [P14] and \emph{``conversations about life and the news''} [P27]. As with prior studies conducted in domestic settings \parencite{harper2017interrogative}, additional topics were often stimulated by reference to objects or other people visible in participants' home environments [P9]. Web links shared during video calls [P19] were also sources of conversational stimulation. Nevertheless, COVID dominated conversation during social talk, alongside day-to-day exchanges such as common hobbies and interests.

\subsubsectionnonum{Conversational Phases}
Our findings echo previous literature \parencite{mirivel2005premeeting} in identifying small talk as a common pre-meeting ritual, with many participants mentioning that video conferencing meetings began with \emph{"niceties"} [P11] before going straight into work topics in addition to \emph{"signing off in an informal way"} [P14] at the end of the meeting. 

\begin{quote} 
\textit{``Either in the beginning or in the end. I mean, it's very rare that something would come up in the middle that does not strictly relate with the project at hand.''}-[P16]
\end{quote}

\begin{quote} 
\textit{``I think really the difference is trying to avoid miscommunication. And this is why it's kind of, you're more inclined to be formal. Except for the beginning and end of a call.''}-[P7]
\end{quote}

Professionalism was maintained throughout meetings, with social interactions limited to the beginning and end (as noted above) and during instances where participants were waiting for others to join the meeting. Although at times social talk occurred during the meeting, the person holding the meeting would \emph{``get straight into business''} [P1] once everyone arrived due to \emph{``limited time''} [P5]. This left the proportion of social talk to be \emph{``around 10 or 15''} [P27] percent of the meeting, as quoted by various participants (P4,P15,P23,P25,P27). It was pointed out that such pre-meeting small talk is not unique to video conferencing:

\begin{quote}
\textit{``people might sort of have like a little kind of. “Hi, how are you”, and you know just same as you would if you went into a meeting room in-person.''}-[P14]
\end{quote}

Yet, participants emphasized the increased importance of this small window of social interaction while working from home, potentially in isolation.

\begin{quote}
\textit{``I look forward to zoom meetings, because I know there's always going to be the opportunity to have a little bit of chit chat conversation with my colleagues. That I'm obviously not having because we're not physically on campus at the moment''
}-[P23]
\end{quote}

\begin{quote}
\textit{``you are completely isolated. You are alone. So, to be honest, it's nice to be able to to have that social aspect.
To - even if it's from a distance - to have some level of normality''}-[P15]
\end{quote}

As such, what was previously a gesture of politeness with in-person meetings appears to have taken on greater significance in WFH scenarios. 

\subsection{Relationships}
In the interviews, participants were asked to discuss their relationship with colleagues. Many reported that they had the closest bonds with colleagues who participated in the study with them.

\begin{quote} 
\textit{``they're both people I would consider friends so they're more than just coworkers, they're not, you know, they're not the closest friends I have but there's certainly...I'm a lot more friendly with them, than I will be with other coworkers. We would be more willing to go out and do something social outside of work as well. When we could actually do that.''}-[P18]
\end{quote}

However, few others mentioned spending time with these colleagues outside of work. Instead, socializing with these close colleagues tended to occur at company events such as Christmas dinners.

\begin{quote} 
\textit{``The only socializing outside of work time we do is at the Christmas party at the Institute or some other things that they do, social events.''}-[P19]
\end{quote}

\begin{quote} 
\textit{``Everyone is friendly but, well people enjoy interaction but it’s very infrequently that those interactions, social communication, extends after work hours.''}-[P26]
\end{quote}

Colleagues' relationships were mainly defined based on frequency of interaction and the opportunities for these interactions provided by their organizations. The time that colleagues had known each other was also a common determinant of friendship. 

\subsubsectionnonum{Conversational Barriers}
The relationship between colleagues also determined the ease of conversation and parallels between their experience of VC compared to in-person conversations. Although awkwardness was frequently mentioned regarding VC, relationships that were forged pre-pandemic determined the level of comfort participants felt conversing with colleagues over video. 

\begin{quote} 
\textit{``I know them very well, I don't feel like there's anything. It doesn't feel difficult. It's easy to talk to them. As easy as in person.''}-[P10]
\end{quote}

\begin{quote} 
\textit{``I think it's probably more difficult in informal situations where you're not familiar with people and there’s maybe some awkwardness around that, where there's no awkwardness at all when you're talking with friends.''}-[P23]
\end{quote}

Despite reporting feeling less awkward when conversing on VC platforms with colleagues they felt closer to, barriers to natural conversations were still prevalent. Participants noted that VC experiences were often frustrating, with some mentioning fatigue. Many also noted difficulties when trying to infer meaning from physical gestures. While this is effortless in in-person encounters, participants felt they had to \emph{"adapt that on to a Zoom"} [P15] in remote interactions. 

\begin{quote} 
\textit{``it's frustrating because I’m missing my colleagues slash friends and it's kind of the Zoom conversations like approximating that, except for it's not for everyone to kind of pretending that it's that it's just a normal chat, but it's all slightly strained and awkward.''}-[P11]
\end{quote}

\begin{quote} 
\textit{``we are kind of friends or, you know, colleagues, and only engaging socially, it’s tiring..''}-[P3]
\end{quote}

Overall, participants mentioned an easier flow in conversations on video conferencing among friends, allowing them to base these mediated interactions on previous in-person conversations. They mentioned that similarities to in-person conversations depended on whether they knew each other well. Colleagues would fall into \emph{``established roles''} [P23] they had in in-person conversations. One colleague mentioned that because she started working shortly before the pandemic, these interactions \emph{``wouldn't be probably much different because we are not that close''} [P3]. 

\subsubsectionnonum{Status and Roles}
Echoing previous literature examining power in the workplace \parencite{fairclough2013critical,fairclough2001language}, participants mentioned status as an important factor that dictated conversational behaviors between colleagues. Although participants that were recruited for the study were in a similar career stage, characteristics such as organizational tenure and age were attributed to seniority, as mentioned in social psychology literature \parencite{montepare1998person}. Participants mentioned these dynamics were magnified in video conferencing environments due to increased possibilities of interrupting colleagues, in addition to the presence of senior members in organized formal social talk events. 

\begin{quote} 
\textit{``even though everyone is very friendly, there's also maybe a hierarchy, you know, head of school and people who are more, much more senior than I am and that also plays into...It's different than with, you know, with your friends outside work [..] I think all the kind of structures around seniority and stuff, maybe gets magnified in the Zoom space because you're being extra careful about not interrupting someone who's trying to say something important..''}~[P10]
\end{quote}

\begin{quote} 
\textit{``I could say with our, you know, again, our coffee afternoons, because it's still social, but it's more formal. You know, [..] typically the department head will be sort of the speaker and other people can speak but it's almost as if, it's as if you have a meeting and chairing a meeting..''}~[P18]
\end{quote}

Formalities were also maintained when engaging in social talk with colleagues, with people often deferring to senior members even in social talk situations.   
\begin{quote} 
\textit{``I do feel as equals. It's pretty easy for me to talk and to lead the conversation. I don't feel scared to do this but if there was someone that I thought of as an authority figure above me then I’d wait for them to speak more..''}~[P1]
\end{quote}

This was often due to uncertainty around when to speak on VC. Indeed, turn-taking issues caused by a perceived lack of structure in these conversations, and unclear roles, were widely mentioned by participants, particularly in unstructured social meetings. In contrast, formal meetings provided clarity due to set agendas and professional roles assigned depending on expertise. 

\begin{quote} 
\textit{``I’ve done a few PhD supervisions and they've been amongst the easiest meetings I have. It’s not because I’m in charge, there’s a very clear hierarchy which is normally something I kind of try and avoid. But (inaudible) but just from a practical perspective, the kind of roles are more clear cut...''}~[P11]
\end{quote}

\begin{quote} 
\textit{``there's always this point in like turn-taking, who gets to talk and when. In particular if it’s not professional but social because you don’t actually have, you're not reporting on projects, you're not taking turns....''}~[P25]
\end{quote}

Interestingly, host privileges also affected the roles participants took on in the conversational session. During the session, one of the three participants was randomly assigned to be the host on Zoom. This was to allow the experimenter to leave the room as participants could chat unsupervised.  
Some participants reported feeling an extra sense of responsibility in moderating the conversation due to having the role of `Host' assigned to them. 

\begin{quote} 
\textit{``I was honored, but I also, I also felt responsibility and so it felt like I should start the conversation at that point [..] even though it wasn't, it wasn't an official role...''}~[P23]
\end{quote}

While present in formal meetings, the role of the moderator was missing in social talk. Although participants dominated the conversation depending on the conversational topic, resulting in dynamic roles, a moderator role was taken on or assigned. Participants mentioned various benefits in assigning a moderator and mentioned moderator qualities such as \emph{``direct[ing] attention appropriately''} [P11] and \emph{``dictating the topic''} [P14]. These individuals would \emph{``jump in to fill silences''} [P23] and \emph{``move the conversation along''} [P30]. Senior members would often take on this role. 

\begin{quote} 
\textit{``I'd say I kind of took on the leader, per se, but just the one that gets the conversations going or bring a new topic...''}~[P10]
\end{quote}

\begin{quote} 
\textit{``I definitely took on the role of mammy\footnote{A colloquialism for `mother'}. Yeah, so guiding the conversation, or at least I felt I did [..] I think that's a function of two things, one being given, being made host, but two, also being more senior to the others...''}~[P29]
\end{quote}

Turn-taking issues on VC, in addition to the influence of professional status, dictated conversational behavior. Responsibility was directed or assumed in order to avoid \emph{awkward} [P10, P30] moments of interruption and simultaneous speech. These issues are due to the technical limitations of the platforms that create barriers to participation, which we elaborate on in the next section. 

\subsection{Etiquette and Politeness}
Inherent constraints of video conferencing as an interaction modality required more conscious and purposeful actions of social etiquette and politeness than participants felt would otherwise have been present in in-person social talk. Although this made conversations feel less natural, the heightened awareness of turn-taking and interruptions was necessary for maintaining remote collegiality.

\subsubsectionnonum{Attention, Multitasking, and Social Acceptability of Home Distractions}
Challenges in maintaining attention during teleconference calls were frequently reported by participants. In particular, they reported difficulty engaging in conversational topics while using video conferencing tools due to engaging in multitasking behavior. Many saw this as a result of the lack of behavioral cues available on these tools. 

\begin{quote} 
\textit{``sometimes people talk about something that I am not that interested, especially in the social part the conversation. I may not concentrate that very well. I may go back to my checking emails or sometimes phones...''}~[P4]
\end{quote}

However, participants also reported that multitasking behavior was acceptable, noting that these behaviors were tolerated due to distractions in the home environment. Multitasking behavior was reported on the same device colleagues used for video conferencing, echoing previous findings from \cite{marlow2016taking}. Such behaviors were seen as more acceptable to those multitasking when they took place on the same screen as a video call. 

\begin{quote} 
\textit{``usually assume that when you're having a conversation with somebody who is just looking at their screen and then nothing else, that they are listening....''}~[P8]
\end{quote}

\begin{quote} 
\textit{``it doesn't matter so much because we can be browsing and reading other things while talking....''}~[P19]
\end{quote}

Some participants mentioned recognizing multitasking behavior among their colleagues, through behaviors such as diverted eye gaze from the camera (as similarly reported by \cite{marlow2016taking}). However, many also reported being unable to recognize multitasking behavior due to the camera position providing illusions of mutual gaze:

\begin{quote} 
\textit{``But on the internet it’s a bit different it’s accepted that they move, they look at a different screen for a second. Or something happens in the room and they just have to look away''} [P6]. 
\end{quote}

\begin{quote} 
\textit{``with video conferencing like with Zoom and things like that, it will be just the same. You know, like, people can be, can be in front of the camera, but looking at a totally different window. So god knows...''}~[P22]
\end{quote}

Indeed, such multitasking behaviors were not necessarily conscious efforts - attentional issues were exacerbated in multiparty settings, particularly when colleagues were not addressed or speaking. Participants noted that in one-on-one conversations, they were able to more easily identify speaking opportunities and discuss intimate topics. 

\begin{quote} 
\textit{``you have more focus but possibly more connection then maybe because like, there aren't any moments where you have to wait until other people communicate and you just listen to them...''}~[P26]
\end{quote}

In summary, acceptance of `zoning out' and being distracted by emails, phones, or general domestic life, sets a casual collegial tone, which in real life is enabled by the ability for conversational participants to enter and leave with no social barriers to doing so. 

\subsubsectionnonum{Paralinguistic Cues}
Turn-taking difficulties were attributed to the limited visual display of body language that is normally used to signal interlocutors' level of engagement, despite experimental studies showing otherwise \parencite{sellen1995remote}. Participants mentioned not being able to read the conversational scene and felt they were missing out on social cues that signal opportunities for participation among colleagues and signal levels of engagement/attention among interlocutors. 
When comparing experiences of chatting with their colleagues in person, participants mentioned not being able to identify speaking opportunities on video-mediated conversations as particularly problematic, especially in larger groups. 

\begin{quote} 
\textit{``you know if someone wants to chime in and say something
I find it much more difficult to read on a meeting over Zoom....''}~[P10]
\end{quote}

\begin{quote} 
\textit{``there's less to do kind of if you're just reading one person and looking at their one face rather than two and you don't have to think about cutting across the third person if it's just the two of you....''}~[P14]
\end{quote}

\begin{quote} 
\textit{``I mean, it's harder when the group is larger than four. I think I miss the body language miss the tone in the face and it becomes more of a round circle....''}~[P9]
\end{quote}

Another way participants mentioned reading attentiveness and engagement among colleagues was through facial expression as similarly reported by \cite{isaacs1994video}. 

\begin{quote} 
\textit{``it is important to have the facial cues and the video on because you do get a judge, you can tell straight away if someone is being quiet on it, if they're distracted on it, in that sense, you buck up and ask them, but if someone's happy enough to be talking away again...''}~[P15]
\end{quote}

\begin{quote} 
\textit{``he doesn't nod he winks and I think that really gives you a lot of confirmation that he's you know he's actually listening and engaged....''}~[P25]
\end{quote}

Although participants also noted backchanneling behavior in both in-person and VC contexts, many mentioned that they \emph{``assume that people are listening''} [P23] and \emph{``didn't need to do any special type of estimating''} [P27] in VC scenarios. In some instances where conversations would extend for a longer period of time (for example, in story telling contexts) participants would then evaluate listening behavior through the follow-up questions and directed gaze of others. 

\begin{quote} 
\textit{``Well, you, you just have to have your good [faith]. And think that they are listening...''}~[P22]
\end{quote}

\begin{quote} 
\textit{``why be there online if you're not listening, right?...''}~[P24]
\end{quote}

\begin{quote} 
\textit{``I was just narrating a specific incident. So, you know, it was. And I found that they were listening so they were engaged. They were looking at me, or the screen and asking questions. So I knew that they were listening. Sometimes if I'm in a longer conversation. I would you know kind of look for cues....''}~[P28]
\end{quote}

It is likely that, due to the prevalence of technical faults such as latency, paying attention to body language becomes paramount to conversations on VC. As some participants noted, the absence of body language often conveyed a technical error: \emph{``people are so still and quiet on Zoom that it looks like their screen has frozen sometimes''} [P23]. Compared to in-person environments, participants reported picking up cues as \emph{``more difficult over Zoom''} [P21] with one reporting fatigue: \emph{``I feel it's a bit tiring because you're constantly, you know, aware of your mimic being analysed and perceived''} [P25]. Although participants only mentioned paying attention to backchanneling in instances where conversations stretched for a longer period (also called conversational chunks), attentiveness is assumed, rather than monitored, by participants when engaging in social talk. 

\subsubsectionnonum{Common Ground and Mutual Responsibility}
As outlined in previous literature \parencite{anderson2006achieving} on multiparty conversations, common ground is more difficult to achieve when three or more people are involved in conversation. Further, factors such as divergences in culture and common knowledge result in speakers having to design utterances to tailor to addressees' perspectives \parencite{yoon2019contextual}. Similarly, in our results, participants reported having to curate conversations to multiple addressees. 

\begin{quote} 
\textit{``if you’re like three people and the two people have something in common, and somebody plays an instrument, and somebody plays an instrument, and the third person is not really into any instrument at all, and I could see that this could be a problem. But this is all related to the people talking and if they understand what they are talking about and if they understand what the other person knows or doesn’t know....''}~[P20]
\end{quote}

\begin{quote} 
\textit{``everything just is very general because people naturally don't want people to be left out or bored. So then the content must be something that has, that speaks to everybody in the room....''}~[P10]
\end{quote}

This was mainly due to the limited affordances VC provides in relation to facilitating side talk or parallel conversations, as was also reported by \cite{buxton1997interfaces}. Participants specifically mentioned audio limitations, suggesting that simultaneous dialogue would result in \emph{``a bunch of noise''} [P14] meaning it was impossible to \emph{``turn to the person next to you and talk to them about something that maybe other people don't care about''} [P10].  

\begin{quote} 
\textit{``It's harder to have a conversation because usually what would happen if we were all together in person, you might splinter off and talk to one person about this specific thing ....''}~[P14]
\end{quote}

As a result, conversations revolved around lighter topics including humor and banter. As a participant noted, conversational topics were catered to an \emph{``all-ages audience''} [P17].

\begin{quote} 
\textit{``I would say that if it's in mixed company you tend to have more lighter subjects...''}~[P17]
\end{quote}

\begin{quote} 
\textit{``It's great to have the group, you know, because then it turns more - that's more about banter and teasing and slagging and, more flippant or frivolous....''}~[P29]
\end{quote}

Politeness demonstrated in VC calls expanded to ensuring that all conversational participants were included in some way. Thus, the value and depth of social talk is strongly influenced by the homogeneity of participants, more so than in real life coffee break chat. During in-person collegial situations, those not directly involved in a conversation can drift away without a formal departure, or join a separate group. Remote collegiality is thus facilitated through careful awareness of existing participants.

\subsubsectionnonum{Interruptions}
Participants frequently mentioned interruptions as a key cause of frustration in their experience of using VC. Interruptions resulted from a number of technical barriers such as latency, limited visual display and feedback, in addition to all conversation being delivered through a single combined audio channel. When discussing interruptions, participants described strategies to reduce their impact, and the consequences of their inevitable occurrence. In addition to these technical issues, excitement and enthusiasm in conversations were also mentioned as a regular source of interruptions. 

\begin{quote} 
\textit{``Try not to interrupt them! And that's why the technical issue is so annoying because you can cut across somebody without meaning to and then nobody else in the room can hear anything....''}~[P10]
\end{quote}

\begin{quote} 
\textit{``sometimes it happens that people interrupt people [..] and sometimes if things get heated or excited, people start speaking at the same time....''}~[P12]
\end{quote}

Moments of silence were described as a common consequence of interruptions. This often happened when conversations became heated or boisterous and simultaneous speech occurred. Silence would then follow as colleagues figured out which speaker would hold the floor.

\begin{quote} 
\textit{``the biggest challenge for me is when two person wants to talk in the same time. As I don't like it. It's a strange situation because afterwards. Everyone stay quiet and it lasts for, for a moment....''}~[P3]
\end{quote}

Colleagues would then adapt strategies to manage these awkward silences, which would again result in simultaneous speech. One participant described the process of deciding who would have the floor as an \emph{``awkward shuffle''} [P18].  

\begin{quote} 
\textit{``It's like when you know when you're walking along the road and nobody's looking the other way towards you. And you both kind of go, you go to pass each other but you go in the same direction and you're kind of going back and forth [..] till somebody eventually kind of steps aside and says, Oh, you first or whatever. It's that kind of thing. But in in conversational form...''}~[P23]
\end{quote}

\begin{quote} 
\textit{``a bit of kind of talking over each other not meaning to and then everyone pausing at the same time and not moving. So it’s kind of the usual Zoom awkwardness...''}~[P11]
\end{quote}

In overcoming this, excessive politeness emerged as a common feature of different strategies among colleagues, though this often exacerbated the matter. For instance, a common approach was to cede the floor: \emph{``My personal strategy is just to let the other person talk. I will say, Oh, no, you go ahead''} [P18]. However, this often resulted in over-deferring: \emph{``if I talk over someone I might say, go ahead and they might say go ahead''} [P19]. These situations were perceived as \emph{``awkward''} [P18] as some colleagues would \emph{``wait and wait and wait''} [P24] until the next speaker was established.

\subsection{Connectedness Between Themes}
Our three major themes individually represent the high-level factors that influence how workplace social talk is conducted through video conferencing. The required scheduling and explicit purpose of conversations, along with the etiquette norms that must be applied with video conferencing chats, are representative of the unique challenges of recreating the ‘water cooler conversation’ remotely. The pre-pandemic relationships of colleagues then serve to ameliorate or exacerbate the challenges of such conversations. We thus identify relationships as the common glue to determine the opportunities and etiquette around social talk in remote work. As our participants were able to draw from in-person experiences, relationships determined readiness to interact with colleagues to a greater extent.  This in turn determined the ease of conversing on VC, due to prior common ground.

However, irrespective of attendee relationships, the unwritten rules of etiquette that arise from VC constraints are a major influence of the content of conversations. In finding the ‘common ground’ necessary for including all participants in a VC chat, this often leads to ‘talking shop’ with work-related conversation acting as the lingua franca of colleagues whose personal interests are increasingly less likely to overlap as the number of participants increases. In a similar vein, the strategies of dealing with the interruptions caused by technical limitations (excessive politeness and tentative taking of the floor) are at odds with the purpose of building social connectedness that participants aim to achieve. As such, the sub-themes identified around politeness and etiquette serve to steer conversations away from their intended purpose.

Conversely, the purpose of a scheduled conversation (e.g., shared lunchtime, departmental coffee morning) determine the perceived need to adhere to some of the referenced politeness conventions, regardless of prior relationships. During informal scheduled commensality, the need to reinforce engagement with paralinguistic cues and minimise distractions was not perceived as so important. In social conversations perceived as more intimate or formal, politeness norms then begin to play a part, and expectations are raised to consider multi-tasking behaviour and disengaged body language as impolite. As such, the theme of conversational purpose influences that of etiquette in online social gatherings much the same way as their in-person equivalents.

\section{Discussion}
Social talk is important to foster workplace relations, helping in facilitating collaboration \parencite{kraut1990informal}, improving social bonds \parencite{coupland2014small} as well as leading to better job satisfaction \parencite{riordan1995opportunity}. The recent COVID-19 pandemic has seen an increased number of people working from home, with many organizations planning to support WFH practices post-pandemic. Consequently, supporting social interactions between groups of colleagues and promoting collegiality remotely becomes a significant challenge for the future of work. Our study uses the unique WFH context afforded by the COVID-19 pandemic to explore how social talk between colleagues is being facilitated through VC to support remote collegiality. Through analysis of semi-structured interviews, to the best of our knowledge, our study is the first to explore the ways in which colleagues facilitate multiparty group social talk through VC when working from home.

Our interviews highlight three key themes relevant to workers' experiences of remote social talk with colleagues. The first theme focused on the changing purposes and opportunities afforded by using VC for social talk with colleagues. Participants mentioned that VC tools reduce physical barriers to social talk that exist in office-based interactions, allowing them to extend the range of social contacts in their organization, as well as allowing them to enhance existing bonds. The start and end of professional meetings, alongside waiting for members to arrive in virtual meetings, were common points where social talk between colleagues occurred. More structured social activities (like virtual quizzes and regular scheduled virtual tea or coffee breaks) were also a major source of social conversation, with workers feeling the need to specifically schedule regular social activities to compensate for the lack of in-person social contact. This increased formality has meant that social talk between groups tends to lack spontaneity, requiring conscious effort and planning, while sometimes being seen as yet another work-oriented activity. 

The second theme showed that workers perceived the nature of existing relationships and status of addressees as significant influences of social talk. Participants reported feeling more comfortable with colleagues they had social relationships with pre-pandemic, allowing them to more easily identify and fall into previously established social roles. The job status of team members within their organization also influenced the dynamics of social talk within groups, especially in planned social occasions. Colleagues were more sensitive about interrupting and correct turn-taking procedures when higher-status members of the organization were present. Being labelled as the host of a social occasion also influenced the status of colleagues in a group call, with the organizer perceiving themselves as the facilitator of the conversation. 

The third theme identified through our interview data highlighted the challenges of mapping etiquette and politeness when interacting socially through VC with multiple partners. Multitasking was seen as acceptable when a member of the group was not speaking, or felt the topic was not relevant to them. Colleagues also emphasized the difficulty in monitoring paralinguistic cues, such as gaze, that signalled a desire to take the floor, or to identify whether team members were engaged in the conversation, with colleagues having to assume that their addressees were attentive. People were also aware of having to curate the content of social talk to multiple addressees, adequately taking into account collective common ground between all group members. This severely constrained the dialogue as there was no way to naturally `splinter off' into side conversations as would be the case with in-person scenarios. The potential to interrupt and speak over each other was also a concern in these multiparty conversations, leading to conversational breakdown. Silences were common as groups collectively figured out who could have the floor following an interruption, which could result in several speakers ceding the floor simultaneously. 

Our understanding of VC social talk between colleagues is thus represented through the three themes separately, as well as the interplay between them. Below we discuss the implications of our findings, focusing on differences in VC and in-person social talk, the role of relationships among colleagues, and the opportunities afforded by VC to create and maintain bonds. We also focus on conversation etiquette as well as the social and technical barriers to VC social talk, exploring design and best practice implications and future areas of research towards supporting multiparty social talk more effectively in VC.

\subsection{Perceived Differences between In-person and VC Social Talk}
Participants mentioned that VC social talk is planned. This contrasts with in-person social talk, which tends to be dyadic, brief, arising spontaneously between interlocutors in close proximity \parencite{kraut1988patterns}, and coming to an end once a third party joins the conversation \parencite{whittaker1994informal}. Our results thus show how VC reduced physical barriers to social talk that exist in office-based interactions. Although they are not indicative of the frequency in which colleagues engage online as opposed to in person, they point to a perceived increase of group size in workplace social talk encounters. However, we also speculate that colleagues may be more opportunistic in including more participants in VC calls, as they attempted to leverage the affordances of these tools during the pandemic.

Past work shows that in-person communication has been purported to be the general means in which informal communication occurs in physical workplaces \parencite{nardi2000interaction,kraut1988patterns}, while in remote work, our results show that VC is the primary means by which social talk occurs. However, we also noted the role of instant messaging, team tools, and email in coordinating these interactions - echoing what \cite{nardi2000interaction} describe as \emph{outeraction}: the process of negotiating the availability of participants for social talk.
This need for negotiation and planning of interactions appears to have prompted the adoption of formal social talk practices, similar to that observed in in-person planned interactions \parencite{kraut1990informal}.

The planned nature of remote social talk also results in decreased interruptions of people's work activities to engage in such talk. In shared workplace settings, social talk is initiated by signalling conversational intent through mutual gaze, oftentimes when a recipient is working \parencite{whittaker1994informal}.
The planning of dedicated social talk time on VC eliminates these sporadic disruptions of work. Despite these interactions being planned in remote work contexts, our participants still reported that getting back to work was a common concern in these situations. This resonates with prior findings on media spaces, where appearing too social in interactions with these technologies are a key concern \parencite{karahalios2009social}, and with findings on remote workers' cognizance of the need to appear productive in remote work environments\parencite{halford2005hybrid}. We speculate that these concerns when conducting social talk through VC will likely persist post-pandemic.

Our participants described that joint activities with their colleagues provided a clearer purpose to scheduled social time. To facilitate conversations around topics outside of COVID, activities such as quizzes and commensality were planned into participants' work days, the latter supporting past work on mealtime awareness systems \parencite{grevet2012eating,o2012food}. Based on our results, we identify that there are opportunities to facilitate social connectedness through shared experiences, which may compensate for a lack of shared physical space in VC settings. We speculate that shared activities in VC social talk are likely to persist as remote work becomes more hybrid, but may need to consider different types of activities so as to facilitate remote and in-person team members. Yet future would need to explore this more fully.

\subsection{Mediating Relationships in VC Social Talk}
While pre-existing relationships mediate the communication that takes place, the mandatory move to VC also influences relationships old and new. As our second theme reflects, chance encounters in communal areas that began relationships in pre-COVID times have been replaced by coffee-break video calls that enable connections with participants in different departments, universities, and (in one group), countries. Overall, our results present new ties forged during the pandemic, contrasting with recent research indicating that networks contracted as colleagues relied on their existing close bonds \parencite{microsofttrends}. 

Although pre-existing bonds between colleagues did not seem to dictate the breadth of their VC social networks, they instead defined media usage patterns, and the \emph{outeraction} that took place between them. For example, messaging platforms such as WhatsApp were used exclusively between colleagues with closer bonds, whereas VC was used by all colleagues for synchronous social communication, irrespective of roles and relationships. Organizations' media infrastructures hence defined the choice of VC applications for social connectedness among their employees when existing ties and \emph{communication zones} were not established. In our results, VC applications were used day-to-day for facilitating collaboration as a result of existing organizational norms, yet these platforms were also adopted for non-task-based interactions.

Our results also point to the increased importance of status in social talk. Although prior research suggests that in-person interlocutors are likely to take on dynamic roles \parencite{gilmartin2018explorations}, our results instead suggest that conversational participants assume roles based on characteristics relating to their profession and seniority. Moreover, host privileges on VC platforms (such as Zoom) prompted participants to assume the role of the \emph{moderator}, with a feeling of responsibility to keep the conversation going. This parallels experiences of in-person dyadic social talk where social talk initiators have more control over a conversation \parencite{nardi2002place}.

\subsection{Etiquette and Common Ground in Multiparty Social Talk}
As acknowledged, formality is a feature of conversational dynamics in task-based VC \parencite{sellen1995remote} and our findings suggest this is also the case in VC-based social talk. Our third theme represented the importance of politeness and etiquette for monitoring turn-taking behavior and ensuring full group inclusion. However, our findings also point to the acceptability of inattention, though this may be an artifact of conversing on VC platforms more generally \parencite{gutwin2002descriptive}. For example, echoing prior work on multitasking in domestic use of VC platforms \parencite{brubaker2012focusing} and teenagers' use of VC for social calls \parencite{suh2018s}, our study reports participants' acceptance of distracted behavior caused by events in the home environment, or simply browsing the internet. We recommend that future research further investigates multitasking acceptability of social talk to delineate conversational, group, or VC qualities that lead to the acceptance of this behavior. 

As well as leading to multitasking, the lack of clear ability to signal inattention or disinterest also seemed to shift the topics covered in multiparty social conversations over VC to be as inclusive as possible. In our interviews, participants mentioned waiting as sub-conversations emerged in group contexts, resulting in inattention. Prior research \parencite{isaacs1994video} on a multiparty VC prototype similarly reports that dyadic conversations would occur around topics of interest, as others waited for the conversation to become more general. Further, findings from task-based multiparty dialogues show that having multiple people involved in a dialogue makes considering common ground when speaking 
more difficult \parencite{anderson2006achieving}, and our work suggests that there is a similar challenge in social talk. Along with a sensitivity to being inclusive to those unfamiliar with the topics being discussed, this seems to lead people to keep topics light to allow for participation, aligning with findings from \cite{gibson2003participation}. We suggest that future work explores the effects of group size and common ground on levels of attention during VC-based social talk.

\subsection{The Importance of Perceived Politeness \& Impression Management}
Despite lab-based experimental studies indicating decreased interruptions and backchannels on VC compared to in-person environments \parencite{sellen1995remote}, our results indicated that this behavior was a pressing concern among colleagues when conversing on VC. However, it is important to keep in mind that in past lab-based studies, participants did not have prior familiarity. In our study, participants had existing workplace relationships, which may explain why interruption behavior was perceived as costly due to the need to manage impressions on these platforms - a concern reported in distributed work arrangements \parencite{leonardi2010connectivity}. 
The need for such `impression management', absent from in-person informal chats with known colleagues, may be explained by the fact that opportunities for social talk are primarily constrained to formalized structured events, like organized social occasions or professional work-related meetings. WFH seems to result in fewer opportunities for frequent spontaneous social talk encounters, which encourage more informal communication behavior shown to influence the transition from formal to informal conversation in prior work \parencite{brown1979speech}. As a suggestion for future work, a longitudinal study could deepen our understanding of how communication patterns in remote collegiality evolve in remote work contexts by explicating the relationship between frequency of interaction and maintaining formality in conversation.

\subsection{Barriers to Fluid Multiparty Social Talk on VC}
Echoing previous work on VC \parencite{buxton1997interfaces,okada1994multiparty,o1993conversations}, issues with latency, the use of a single combined audio channel, and the difficulty in inferring facial expression and body language, were frequently reported by our participants. Limited social cues that signal opportunities to participate in social talk were seen as particularly problematic when interpreting meaning, with colleagues relying on previous in-person encounters to aid them in this regard. 
In terms of limited cues, prior literature \parencite{fish1992evaluating} has argued that the advantages of VC in social talk centre on the use of visual channels. However, limited display sizes in multiparty VC may lead to significant difficulties inferring body language and other subtle visual cues used to guide conversation. This may be exacerbated by the multiparty nature of the situation itself as people may not be able to fully attend to all video channels, potentially increasing cognitive load when monitoring cues in multiparty settings \parencite{jackson2000impact}. This difficulty in identifying visual cues may also add to pre-existing challenges in predicting turn-taking alternation patterns in multiparty talk \parencite{wang2010groups}. 

Likewise, technical constraints of VC applications mean that some core aspects of multiparty social talk are not supported. For instance, side conversations, where interlocutors dynamically break off into smaller groups to converse when in multiparty conversations \parencite{parker1988speaking}, are currently not widely supported on VC. These types of conversations are common during in-person interaction \parencite{parker1988speaking}, and provide more dynamic conversations, though reduce significantly as group size increases \parencite{fay2000group}. Although many tools include breakout functionality, this is usually controlled by the host and lacks the spontaneity of in-person social side conversations. When supporting multiparty social talk, VC applications should look to increase the ability for cues to be more readily interpretable, as well as develop methods to facilitate more spontaneous side conversations within multiparty dialogue. Any such solutions should also be sensitive to supporting expressiveness, frequency, and interactivity \parencite{fish1992evaluating}. 

\subsection{Implications for Design}
Our results highlight key challenges and opportunities to better facilitate social talk in VC among remote colleagues. Based on recent analysis of future work trends \parencite{microsofttrends}, it is likely that VC will play a key role in developing remote collegiality between team members. Based on our findings, we outline some key design implications to consider in future VC design for facilitating social talk among colleagues. Our implications consider how connectedness between co-workers can be facilitated within the boundaries of work, without intruding on home life.

\subsubsection{Tools to support planning}
Our work shows that, compared to in-person conversation, VC-based social talk requires significantly more planning, which participants acknowledged as an inevitable consequence of WFH. Despite the apparent irony of scheduling that which embodied spontaneity in the workplace, we suggest that efforts should be expended in supporting scheduled social interactions. Future VC systems should consider functionality that supports the coordination, planning and facilitation of social interactions. Participants mentioned currently switching between tools such as email clients and other instant messaging platforms when planning social VC calls. Echoing previous work \parencite{tang1994supporting}, a tighter coupling of asynchronous and synchronous communication tools would allow remote informal conversations to be organized with less effort. By creating tools that support work socialization processes, these interactions may be able to situate themselves in remote colleagues' work days and better facilitate social talk in existing communication tools. Particularly as remote workers move towards task management tools to avoid interruptions in their work day. Planning these interactions in opportune moments during breaks and transitions between different types of tasks may hence provide remote workers with social talk opportunities without disrupting them in focused states. This may lead to better managed interruptions during work, a key challenge to providing a productive and happy workplace \parencite{kaur2020optimizing}. Embedding or linking across these tools within VC platforms would reduce the effort in planning and co-ordination, and thus may lead to social calls being organized more frequently between colleagues. Indeed, a more automated approach wherein VC-based bots support or initiate social VC interactions across these channels, may also relieve the burden of planning and initiation highlighted by participants. 

\subsubsection{Supporting `playful' social interactions}
Similar to the design of engaging spaces in the office that would allow colleagues to easily initiate conversations out of professional roles \parencite {gallacher2015mood}, we suggest that VC tools should provide environments and features by integrating \emph{playfulness} \parencite{salen2004rules}; in social talk applications. In VC, design practitioners can support less formal and more playful ways of expressing opinions and reactions, as well as playful ways to take the floor, to improve social talk experiences (e.g., Microsoft Teams Together Mode \footnote{https://news.microsoft.com/innovation-stories/microsoft-teams-together-mode/}, Ohyay \footnote{https://ohyay.co/}, Mozilla Hubs \footnote{https://hubs.mozilla.com/}, Gather Town \footnote{https://www.gather.town/}). Our participants expressed frustration at the structured turn-taking and awkwardness in resolving interruptions in VC. Acknowledging the constraints of VC, including a single combined audio channel, designers could facilitate interactions through non-verbal communication devices such as emojis or augmented reality interactions (such as video-filters to signal emotional reactions) to encourage colleagues to participate without fear of interrupting others. We see efforts towards integrating this in existing VC platforms, such as the use of emojis to signal reactions or the desire to talk in Microsoft Teams and Zoom. Although work is needed to explore the impact of this functionality in social talk, it may promote participation among inactive interlocutors, encouraging them to add to the conversation, especially when there is a status difference between interlocutors. 

What is more, having VC platforms support or supply games or shared activities, similar to what has been proposed in prior work on VC in the home \parencite{follmer2010video,hunter2014waazam}, may also allow for more effective social interactions. Our work highlights that social talk with colleagues tends to be most effective when focused on a shared activity. Embedding quiz bots (e.g., Slack's Polly \footnote{https://www.polly.ai/slack-poll}) or other games would allow for shared social activities to be engaged with as part of a social VC call, or as part of a more social section of a task-focused VC meeting. It would also reduce the effort needed to organize and prepare such activities, thus making these types of interactions more effortless. Finally, including group activity features may help colleagues dissociate VC tools from their professional task-oriented context of use, allowing them to clearly distinguish `work talk' from `social talk' time on VC.

\subsubsection{De-emphasizing status \& encouraging participation in social talk}
As our results suggest, social talk in VC between colleagues is still sometimes seen as work-oriented, with status influencing who organizes and how colleagues participate in social conversations. In particular, senior members tend to take on the role of the \emph{moderator}, taking responsibility for organizing social conversations and filling gaps in conversation. Status of team members also seems to influence how they engage in interaction, with those in a junior role experiencing heightened awareness of potentially interrupting senior team members. Design practitioners should look to de-emphasize these status effects to improve social VC experiences. For instance, rather than senior team members instigating social VC interactions, bots could be used to act as \emph{social catalysts} \parencite{karahalios2009social}, reducing the burden on senior team members to initiate social talk. They could be used to initiate activities during VC interactions or lead social meeting planning. This could be complemented by more formal communications based training for staff on norms and processes when instigating social talk remotely. Such initiatives could lead to social VC interactions being seen less as the responsibility of senior team members, removing status issues in social talk execution and planning. Indeed, specific functionality could also be added to VC platforms to encourage and equalize participation and engagement in social talk across teams. Similar to recent work on assessing meeting dynamics \parencite{meeetingcoach} and spotlighting audience reactions in VC meetings \parencite{affect_spot}, a bot could be embedded in social talk VC sessions to read conversational dynamics or highlight specific reactions so as to encourage more social talk and support those that have not contributed within a session. This type of functionality could be useful to highlight opportunities for engagement in group conversations and encourage a wider range of contributions from group members.

\subsubsection{Remote Commensality in VC Social Talk}
In a similar vein, we suggest designers explore how to incorporate functionality that integrates daily routines into future social talk VC systems. As our findings suggest, colleagues found ways to connect through commensality, planning these experiences into their day to replicate their in-person equivalents. As colleagues' mealtimes are more likely to co-occur \parencite{grevet2012eating}, adopting the design of mealtime systems would allow for colleagues to move away from scheduling social talk, allowing them to enter and leave commensality experiences as they finish their meal. Although we acknowledge that colleagues may choose to spend their mealtimes with family or those they cohabit with to take advantage of WFH arrangements, creating organizational opportunities to socialise may allow new hires and colleagues who may rely more on workplace relationships to initiate conversations around joint contexts. To support these experiences, we suggest that VC or team messaging systems could prompt users to participate in lunch or tea breaks in pre-defined time periods through the day. By doing so, colleagues can engage and disengage in conversations as they eat, moving in and out of conversations \parencite{grevet2012eating} This would further provide colleagues with more formalised socialising routines, mitigating concerns relating to productivity, allowing boundaries to be managed (an issue in remote work \parencite{kossek2016managing}) and making social talk less interruptive in employees' workflows. In line with prior literature on remote commensality in the home, we suggest that food interaction strategies be integrated into VC systems, by creating video filters (e.g., virtual food items) that provide the illusion of being together in a space \parencite{foley2010netpot} or by introducing avatars in these spaces to support these shared experiences \parencite{karahalios2005chit}. We urge future research to explore these ideas in the workplace context, since current applications and findings have focused on experiences of friends and family.

\subsection{Best Practices for Planned Remote Social Talk}

Our work also has implications for best practice when initiating and engaging in remote social talk between colleagues, as we move towards a post-pandemic workplace where hybrid and remote teams are likely to be commonplace. We outline some key guidelines below.

\subsubsection{Leverage rituals and activities for social talk}
Our study data highlights that, much like in in-person interaction between colleagues \parencite{mirivel2005premeeting}, social talk tends to occur at the start and end of more formal VC meetings. We also see that VC-based social talk is prevalent when engaged in shared activities or when organized around existing daily practices (e.g. coffee breaks, lunch). We therefore suggest that in order to foster remote collegiality, social talk organisers should focus on leveraging these existing rituals. For instance, meetings should explicitly allocate time for this type of interaction to unfold. Creating rituals whereby a certain time slot before formal meetings is used for social talk, either in a breakout session or in the virtual meeting room itself, may encourage colleagues to converse out of professional roles and lead to better meeting outcomes. Additionally, reframing social time as a non-intrusive portion of the working day could alleviate feelings of anxiety to return to task-focused time. People looking at ways of engaging teams in social talk could also keep VC-based meet-ups to times that link with rituals like lunch and coffee breaks, whilst also using activities such as quizzes to encourage participants to engage socially with others.

\subsubsection{Be accepting of multitasking}
Our results also suggest that multitasking during social talk is common, being accepted from both a speaker and listener point of view. Participants reported difficulties sustaining attention in social talk, especially when topics were not personally relevant. This multitasking is also likely to increase as meetings stretch for longer periods \parencite{cao2021large}. With hybrid and fully remote teams certain to remain an integral part of organizations for the foreseeable future, managers and team members should be accepting of this behavior when planning and conducting VC-based social talk. In doing so, this will reduce the cognitive load inherent in both social and professional VC calls and thereby unnecessary `Zoom fatigue' \parencite{wiederhold2020connecting}.

\subsubsection{Open up to colleagues across and between organizations}
As our results have shown, socializing through VC has provided opportunities for people to engage with a wider group of colleagues, overcoming barriers to socializing caused by not sharing a physical workplace. Although physical proximity is paramount in enabling colleagues to converse socially with others \parencite{kraut1988patterns}, our participants' experiences highlight the potential of VC to allow colleagues to socially engage with others outside their immediate team. Clearly, by bridging distances through VC, opportunities can be provided for multi-disciplinary/multi-departmental collaboration within organizations. Moreover, managers could also support networks across organizations by creating a shared social context for individuals to engage with one another, which prior work has shown to facilitate knowledge transfer \parencite{nonaka1998concept}. For people initiating social meetings, we recommend that they engage colleagues in other departments and organizations by creating a shared virtual social space to foster collaboration networks through social talk across and within organizations more widely.

\subsubsection{Consider group size}
Although we suggest that VC allows opportunities for remote colleagiality across teams, organizers of social VC activities should also consider how group size may affect social talk. Using VC to socialize with colleagues while working from home has allowed people to increase the number of people they can converse with socially. Yet, both technical and social challenges can arise when conducting multiparty social talk through VC. For instance, we find that moments when colleagues were heated or boisterous can result in uncertainty over who has the next turn. The use of a single audio channel also makes it more difficult to support in-person dynamics in social multiparty VC interactions, such as supporting side conversations as groups grow in size. Involving large groups may also lead to more general topics of conversation as it becomes more difficult for participants to take common ground into account effectively when considering how to contribute to the conversation. Although there are benefits to VC social talk, issues in facilitating interactivity on VC are clearly prevalent, echoing prior work \parencite{fish1990videowindow}. We therefore suggest that when organizing remote team social conversations, organizers should be aware of these limitations and how the size of the group may affect conversational experience.

\subsection{Limitations}
For our study, we recruited triads from a number of universities. Although this may impact the generalizability of our findings, limiting our results to work practices in research institutes, prior research \parencite{kraut1988patterns} in informal communication has similarly examined these practices among researchers. We believe this demographic provides insights to those with international collaboration networks, as oftentimes researchers with closer topic areas are not situated among their peers in a shared physical space.

So as to help with recruiting coworkers that had worked together before the pandemic, we asked participants to recruit two other colleagues to take part in the conversational session. It is likely that they have a prior relationship with these colleagues, such that our interview findings are therefore likely to be more applicable to those that have a previous relationship (whether weak or strong) rather than those that have no previous ties with their colleagues. Future work should aim to explore the experience of VC-based social talk in those cohorts in more detail.

Our overview provides opportunities and challenges of remote social talk as colleagues work from home. However, our results have not delved further into how social talk fits into existing sets of routines that support social interactions in the home. We urge future research to explore the patterns of social interaction among colleagues and those in the home to better outline how technologies such as VC can support and enhance collegial relationships without creating an additional burden for socializing in WFH environments.

Although our findings are relevant to social talk in VC for remote colleagues more widely, it is important to note that this study was conducted during a pandemic. Even though this provided us with an opportunity to study social talk using VC, as many employees were forced to work from home, the unprecedented manner in which the pandemic has altered daily work practices may also impact how our results represent typical remote working experiences. Moreover, the period during which the study was conducted may also influence our results, as communication patterns might have differed in certain periods during the pandemic, as suggested by prior research \parencite{microsofttrends}. Hence, it is important to contextualize our results in the time period in which the data was collected (August and September 2020). As we move towards a post-pandemic workplace, we feel that the insights from our study can still be used to inform future work as companies look to support fully remote or hybrid working. Comparing more directly the experiences of in-person and partially remote workers to those of the fully remote participants in this study could also draw further insights into aspects of collegiality that are exclusive to fully remote WFH environments.

\section{Conclusion}
VC applications are becoming increasingly integrated into work practices, catering for both formal and informal communication. In our study, we outline key themes in relation to how colleagues maintain collegial ties and factors that influence social talk on VC while working remotely. Our findings demonstrate opportunities VC provides for collegial relationships such as deepening bonds, reducing barriers to connect socially, and maintaining day-to-day social interactions with colleagues. We also report additional challenges in multiparty social talk through VC. We report behaviors that emerge as a result of limitations of VC due to technical barriers and how they result in interruptions, simultaneous talk, and inattention. Our work highlights the opportunities afforded by using VC to facilitate social talk between remote colleagues, in addition to challenges presented in the use of these tools to facilitate multiparty social talk.We outline design implications, and best practices to facilitate social connectedness in distributed teams for managers, providing the foundations for future work on VC social talk systems. 

\articlebodyend

\raggedright
\titleformat*{\section}{\bfseries\Large\centering\MakeUppercase}
\printbibliography

@book{fairclough2001language,
title={Language and power},
author={Fairclough, Norman},
year={2001},
publisher={Harlow : Pearson Professional Education},
edition={2}
}

@book{fairclough2013critical,
title={Critical discourse analysis: The critical study of language},
author={Fairclough, Norman},
year={2013},
publisher={Routledge},
address={New York},
edition={2}
}

@inbook{montepare1998person,
title={Person perception comes of age: The salience and significance of age in social judgments},
author={Montepare, Joann M and Zebrowitz, Leslie A},
booktitle={Advances in experimental social psychology},
volume={30},
pages={93--161},
year={1998},
editor = {Mark P. Zanna},
publisher={Academic Press}
}

@inbook{nardi2002place,
 author={Nardi, Bonnie A and Whittaker, Steve},
 chapter={4},
 booktitle={Distributed Work}, 
 editor={Pamela Hinds and Sara Kiesler},
 title={The Place of Face-to-Face Communication in Distributed Work}, 
 year={2002},
 pages={83-111},
 publisher={The MIT Press}
}

@inproceedings{whittaker1994informal,
title={Informal workplace communication: What is it like and how might we support it?},
author={Whittaker, Steve and Frohlich, David and Daly-Jones, Owen},
booktitle={Proceedings of CHI'94 the SIGCHI conference on Human factors in computing systems},
pages={131--137},
year={1994},
publisher={ACM Press, New York},
url = {https://doi-org.ucd.idm.oclc.org/10.1145/191666.191726},
doi = {10.1145/191666.191726}
}

@article{whittaker1997telenotes,
title={TeleNotes: managing lightweight interactions in the desktop},
author={Whittaker, Steve and Swanson, Jerry and Kucan, Jakov and Sidner, Candy},
journal={ACM Transactions on Computer-Human Interaction (TOCHI)},
volume={4},
pages={137--168},
year={1997},
publisher={ACM New York, NY, USA}
}

@inproceedings{cohen1982speaker,
title={Speaker interaction: Video teleconferences versus face-to-face meetings},
author={Cohen, Karen M},
journal={Proceedings of Teleconferencing and Electronic Communications},
pages={189--199},
year={1982}
}

@article{sellen1995remote,
title={Remote conversations: The effects of mediating talk with technology},
author={Sellen, Abigail J},
journal={Human-computer interaction},
volume={10},
number={4},
pages={401--444},
year={1995},
publisher={Taylor \& Francis}
}

@article{yoon2019contextual,
title={Contextual Integration in Multiparty Audience Design},
author={Yoon, Si On and Brown-Schmidt, Sarah},
journal={Cognitive Science},
volume={43},
number={12},
year={2019},
publisher={Wiley Online Library}
}

@inbook{buxton1997interfaces,
title = {Interfaces for Multiparty Videoconferences},
author = {Buxton, W. and Sellen, Abigail and Sheasby, M.},
editor = {Kathleen Finn and Abigail Sellen and Sylvia Wilbur},
booktitle = {Video-mediated Communication},
year = {1997},
pages = {385–400},
publisher = {Lawrence Erlbaum Associates},
address={NJ}
}

@inproceedings{brubaker2012focusing,
author = {Brubaker, Jed R. and Venolia, Gina and Tang, John C.},
title = {Focusing on Shared Experiences: Moving beyond the Camera in Video Communication},
year = {2012},
isbn = {9781450312103},
publisher = {Association for Computing Machinery},
address = {New York, NY, USA},
url = {https://doi-org.ucd.idm.oclc.org/10.1145/2317956.2317973},
doi = {10.1145/2317956.2317973},
booktitle = {Proceedings of the 2012 Designing Interactive Systems Conference},
pages = {96–105},
numpages = {10},
location = {Newcastle Upon Tyne, United Kingdom},
series = {DIS '12}
}

@inproceedings{marlow2016taking,
author = {Marlow, Jennifer and van Everdingen, Eveline and Avrahami, Daniel},
title = {Taking Notes or Playing Games? Understanding Multitasking in Video Communication},
year = {2016},
isbn = {9781450335928},
publisher = {Association for Computing Machinery},
address = {New York, NY, USA},
url = {https://doi-org.ucd.idm.oclc.org/10.1145/2818048.2819975},
doi = {10.1145/2818048.2819975},
booktitle = {Proceedings of the 19th ACM Conference on Computer-Supported Cooperative Work \& Social Computing},
pages = {1726–1737},
numpages = {12},
location = {San Francisco, California, USA},
series = {CSCW '16}
}

@inproceedings{fish1990videowindow,
title={The VideoWindow system in informal communication},
author={Fish, Robert S and Kraut, Robert E and Chalfonte, Barbara L},
booktitle={Proceedings of the 1990 ACM conference on Computer-supported cooperative work},
pages={1--11},
year={1990},
url = {https://doi-org.ucd.idm.oclc.org/10.1145/99332.99335},
doi = {10.1145/99332.99335}
}

@inproceedings{judge2010sharing,
title={Sharing conversation and sharing life: video conferencing in the home},
author={Judge, Tejinder K and Neustaedter, Carman},
booktitle={Proceedings of the 2010 SIGCHI Conference on Human Factors in Computing Systems},
pages={655--658},
year={2010},
url = {https://doi-org.ucd.idm.oclc.org/10.1145/1753326.1753422},
doi = {10.1145/1753326.1753422}
}

@inproceedings{nardi2000interaction,
title={Interaction and outeraction: instant messaging in action},
author={Nardi, Bonnie A and Whittaker, Steve and Bradner, Erin},
booktitle={Proceedings of the 2000 ACM conference on Computer supported cooperative work},
pages={79--88},
year={2000},
url = {https://doi-org.ucd.idm.oclc.org/10.1145/358916.358975},
doi = {10.1145/358916.358975}
}

@article{mirivel2005premeeting,
title={Premeeting talk: An organizationally crucial form of talk},
author={Mirivel, Julien C and Tracy, Karen},
journal={Research on Language and Social Interaction},
volume={38},
number={1},
pages={1--34},
year={2005},
publisher={Taylor \& Francis}
}

@article{jett2003work,
title={Work interrupted: A closer look at the role of interruptions in organizational life},
author={Jett, Quintus R and George, Jennifer M},
journal={Academy of management Review},
volume={28},
pages={494--507},
year={2003}
}

@article{knutson1986exploration,
title={AN EXPLORATION OF THE FUNCTION OF SMALL TALK IN FRIENDSHIP RELATIONSHIPS.},
author={Knutson, Peggy and Ayers, Joe},
journal={Journal of the Northwest Communication Association},
volume={14},
year={1986},
pages={4--8}
}

@article{brass1984being,
title={Being in the right place: A structural analysis of individual influence in an organization},
author={Brass, Daniel J},
journal={Administrative science quarterly},
volume={29},
number={4},
pages={518--539},
year={1984},
publisher={JSTOR}
}

@article{george1990understanding,
title={Understanding prosocial behavior, sales performance, and turnover: A group-level analysis in a service context.},
author={George, Jennifer M and Bettenhausen, Kenneth},
journal={Journal of applied psychology},
volume={75},
number={6},
pages={698},
year={1990},
publisher={American Psychological Association}
}

@inbook{kraut1990informal,
title={Informal communication in organizations: Form, function, and technology},
author={Kraut, Robert E and Fish, Robert S and Root, Robert W and Chalfonte, Barbara L},
booktitle={Human reactions to technology: Claremont Symposium on Applied Social Psychology},
pages={145--199},
year={1990},
editor={Stuart Oskamp and Shirlynn Spacapan},
publisher={Sage Publications}
}

@article{riordan1995opportunity,
title={The opportunity for friendship in the workplace: An underexplored construct},
author={Riordan, Christine M and Griffeth, Rodger W},
journal={Journal of business and psychology},
volume={10},
pages={141--154},
year={1995},
publisher={Springer}
}

@inproceedings{macik2006virtual,
title={Virtual work: Loneliness, isolation and health outcomes},
author={Macik-Frey, M},
journal={Academy of Management Meeting, Atlanta, Georgia, United States},
note={Paper Presentation},
year={2006},
month={8},
location={Atlanta, Georgia}
}

@article{whittaker1995rethinking,
title={Rethinking video as a technology for interpersonal communications: theory and design implications},
author={Whittaker, Steve},
journal={International Journal of Human-Computer Studies},
volume={42},
number={5},
pages={501--529},
year={1995},
publisher={Elsevier}
}

@inbook{tang2014techniques,
author={Tang, John C and Junuzovic, Sasa and Inkpen, Kori and Venolia, Gina},
title={Techniques for Studying Actual Use of Personal Communication Prototypes},
booktitle={Studying and Designing Technology for Domestic Life},
chapter={11},
pages={207},
publisher={Morgan Kaufmann},
year={2014},
}

@article{leonardi2010connectivity,
title={The connectivity paradox: Using technology to both decrease and increase perceptions of distance in distributed work arrangements},
author={Leonardi, Paul M and Treem, Jeffrey W and Jackson, Michele H},
journal={Journal of Applied Communication Research},
volume={38},
number={1},
pages={85--105},
year={2010},
publisher={Taylor \& Francis Group}
}

@article{harper2017interrogative,
title={The ‘Interrogative Gaze’: Making video calling and messaging ‘accountable’},
author={Harper, Richard and Rintel, Sean and Watson, Rod and O’Hara, Kenton},
journal={Pragmatics},
volume={27},
number={3},
pages={319--350},
year={2017},
publisher={John Benjamins}
}

@article{sproull1984nature,
title={The nature of managerial attention},
author={Sproull, Lee S},
journal={Advances in information processing in organizations},
volume={1},
pages={9--27},
year={1984}
}

@article{coupland1992how,
author = {Justine Coupland and Nikolas Coupland and Jeffrey D. Robinson},
journal = {Language in Society},
number = {2},
pages = {207--230},
publisher = {Cambridge University Press},
title = {{``How are you?'': negotiating phatic communion}},
volume = {21},
year = {1992}
}

@article{lin2015potential,
title={Potential job facilitation benefits of “water cooler” conversations: the importance of social interactions in the workplace},
author={Lin, Iris Y and Kwantes, Catherine T},
journal={The Journal of psychology},
volume={149},
number={3},
pages={239--262},
year={2015},
publisher={Taylor \& Francis}
}

@article{allen2014linking,
title={Linking pre-meeting communication to meeting effectiveness},
author={Allen, Joseph A and Lehmann-Willenbrock, Nale and Landowski, Nicole},
journal={Journal of Managerial Psychology},
volume={29},
number={8},
year={2014},
pages={1064--1081},
publisher={Emerald Group Publishing Limited}
}

@inproceedings{kraut1988patterns,
title={Patterns of contact and communication in scientific research collaboration},
author={Kraut, Robert and Egido, Carmen and Galegher, Jolene},
booktitle={Proceedings of the 1988 ACM conference on Computer-supported cooperative work},
pages={1--12},
year={1988}
}

@inproceedings{sellen1992speech,
title={Speech patterns in video-mediated conversations},
author={Sellen, Abigail J},
booktitle={Proceedings of the SIGCHI conference on Human factors in computing systems},
pages={49--59},
year={1992}
}

@article{o1993conversations,
author = {O'Conaill, Brid and Whittaker, Steve and Wilbur, Sylvia},
title = {Conversations over Video Conferences: An Evaluation of the Spoken Aspects of Video-Mediated Communication},
year = {1993},
issue_date = {December 1993},
publisher = {L. Erlbaum Associates Inc.},
address = {USA},
volume = {8},
number = {4},
issn = {0737-0024},
journal = {Human-Computer Interaction},
pages = {389–428},
numpages = {40}
}

@article{ibarra1992homophily,
author = {Herminia Ibarra},
journal = {Administrative Science Quarterly},
number = {3},
pages = {422--447},
publisher = {[Sage Publications, Inc., Johnson Graduate School of Management, Cornell University]},
title = {Homophily and Differential Returns: Sex Differences in Network Structure and Access in an Advertising Firm},
volume = {37},
year = {1992}
}

@inproceedings{jackson2000impact,
title={Impact of video frame rate on communicative behaviour in two and four party groups},
author={Jackson, Matthew and Anderson, Anne H and McEwan, Rachel and Mullin, Jim},
booktitle={Proceedings of the 2000 ACM conference on Computer supported cooperative work},
pages={11--20},
year={2000},
url = {https://doi-org.ucd.idm.oclc.org/10.1145/358916.358945},
doi = {10.1145/358916.358945}
}

@article{anderson2006achieving,
title={Achieving understanding in face-to-face and video-mediated multiparty interactions},
author={Anderson, Anne H},
journal={Discourse processes},
volume={41},
number={3},
pages={251--287},
year={2006},
publisher={Taylor \& Francis}
}

@article{rintel2013video,
title={Video calling in long-distance relationships: The opportunistic use of audio/video distortions as a relational resource},
author={Rintel, Sean},
journal={The Electronic Journal of Communication/La Revue Electronic de Communication (EJC/REC)},
volume={23},
year={2013}
}

@inproceedings{forghani2014routines,
title={The routines and needs of grandparents and parents for grandparent-grandchild conversations over distance},
author={Forghani, Azadeh and Neustaedter, Carman},
booktitle={Proceedings of the 2014 SIGCHI Conference on Human Factors in Computing Systems},
pages={4177--4186},
year={2014},
url = {https://doi-org.ucd.idm.oclc.org/10.1145/2556288.2557255},
doi = {10.1145/2556288.2557255}
}

@inproceedings{suh2018s,
title={"{I}t's Kind of Boring Looking at Just the Face": How Teens Multitask During Mobile Videochat},
author={Suh, Minhyang and Bentley, Frank and Lottridge, Danielle},
journal={Proceedings of the ACM Conference on Human-Computer Interaction},
volume={2},
number={CSCW},
pages={1--23},
year={2018},
publisher={ACM New York, NY, USA}
}

@article{tews2013does,
title={Does fun pay? The impact of workplace fun on employee turnover and performance},
author={Tews, Michael J and Michel, John W and Stafford, Kathryn},
journal={Cornell Hospitality Quarterly},
volume={54},
number={4},
pages={370--382},
year={2013},
publisher={Sage Publications Sage CA: Los Angeles, CA}
}

@inproceedings{gilmartin2017exploring,
title={Exploring multiparty casual talk for social human-machine dialogue},
author={Gilmartin, Emer and Cowan, Benjamin R and Vogel, Carl and Campbell, Nick},
booktitle={International Conference on Speech and Computer},
pages={370--378},
year={2017},
organization={Springer}
}

@article{gilmartin2018explorations,
title={Explorations in multiparty casual social talk and its relevance for social human machine dialogue},
author={Gilmartin, Emer and Cowan, Benjamin R and Vogel, Carl and Campbell, Nick},
journal={Journal on Multimodal User Interfaces},
volume={12},
number={4},
pages={297--308},
year={2018},
publisher={Springer}
}

@article{nejati2016implications,
title={The implications of high-quality staff break areas for nurses’ health, performance, job satisfaction and retention},
author={Nejati, Adeleh and Rodiek, Susan and Shepley, Mardelle},
journal={Journal of nursing management},
volume={24},
number={4},
pages={512--523},
year={2016},
publisher={Wiley Online Library}
}

@article{isaacs1994video,
title={What video can and cannot do for collaboration: a case study},
author={Isaacs, Ellen A and Tang, John C},
journal={Multimedia systems},
volume={2},
number={2},
pages={63--73},
year={1994},
publisher={Springer}
}

@inproceedings{okada1994multiparty,
author = {Okada, Ken-Ichi and Maeda, Fumihiko and Ichikawaa, Yusuke and Matsushita, Yutaka},
title = {Multiparty Videoconferencing at Virtual Social Distance: MAJIC Design},
year = {1994},
isbn = {0897916891},
publisher = {Association for Computing Machinery},
address = {New York, NY, USA},
url = {https://doi-org.ucd.idm.oclc.org/10.1145/192844.193054},
doi = {10.1145/192844.193054},
booktitle = {Proceedings of the 1994 ACM Conference on Computer Supported Cooperative Work},
pages = {385–393},
numpages = {9},
location = {Chapel Hill, North Carolina, USA},
series = {CSCW '94}
}

@inbook{brown1979speech,
title={Speech as a marker of situation},
editor={K. R. Scherer and H. Giles},
author={Brown, Penelope and Fraser, Colin},
booktitle={Social markers in speech},
pages={33--62},
year={1979},
publisher={Cambridge University Press}
}

@inproceedings{fish1992evaluating,
title={Evaluating video as a technology for informal communication},
author={Fish, Robert S and Kraut, Robert E and Root, Robert W and Rice, Ronald E},
booktitle={Proceedings of the 1992 SIGCHI conference on Human factors in computing systems},
pages={37--48},
year={1992}
}

@article{fay2000group,
title={Group discussion as interactive dialogue or as serial monologue: The influence of group size},
author={Fay, Nicolas and Garrod, Simon and Carletta, Jean},
journal={Psychological science},
volume={11},
number={6},
pages={481--486},
year={2000},
publisher={SAGE Publications Sage CA: Los Angeles, CA}
}

@article{parker1988speaking,
title={Speaking turns in small group interaction: A context-sensitive event sequence model.},
author={Parker, Kevin C},
journal={Journal of Personality and Social Psychology},
volume={54},
number={6},
pages={965},
year={1988},
publisher={American Psychological Association}
}

@inproceedings{wang2010groups,
title={Groups in groups: Conversational similarity in online multicultural multiparty brainstorming},
author={Wang, Hao-Chuan and Fussell, Susan},
booktitle={Proceedings of the 2010 ACM conference on Computer supported cooperative work},
pages={351--360},
year={2010},
url = {https://doi-org.ucd.idm.oclc.org/10.1145/1718918.1718980},
doi = {10.1145/1718918.1718980}
}

@inproceedings{rintel2007maximizing,
author = {Rintel, Sean},
title = {Maximizing Environmental Validity: Remote Recording of Desktop Videoconferencing},
booktitle={Proceedings of the 12th international conference on Human-computer interaction: interaction design and usability},
year = {2007},
isbn = {9783540731047},
publisher = {Springer-Verlag},
address = {Berlin, Heidelberg},
pages = {911–920},
numpages = {10},
location = {Beijing, China},
series = {HCI'07},
doi={10.1007/978-3-540-73105-4_100}
}

@article{gibson2003participation,
title={Participation shifts: Order and differentiation in group conversation},
author={Gibson, David R},
journal={Social forces},
volume={81},
number={4},
pages={1335--1380},
year={2003},
publisher={Oxford University Press}
}

@article{sacks1978simplest,
author = {Harvey Sacks and Emanuel A. Schegloff and Gail Jefferson},
journal = {Language},
number = {4},
pages = {696--735},
publisher = {Linguistic Society of America},
title = {A Simplest Systematics for the Organization of Turn-Taking for Conversation},
volume = {50},
year = {1974}
}

@book{heritage2010talk,
title={Talk-in-action: Identities, interaction and institutions},
author={Heritage, John and Clayman, Steven},
publisher={Malden, MA: Wiley-Blackwell},
year={2010}
}

@book{ford2008women,
title={Women speaking up: Getting and Using Turns in Workplace},
author={Ford, Cecilia E},
year={2008},
publisher={Palgrave Macmillan UK}
}

@inbook{karahalios2009social,
title={Social catalysts for creating sociable media spaces},
author={Karahalios, Karrie G},
booktitle={Media Space 20+ Years of Mediated Life},
pages={75--95},
year={2009},
publisher={Springer},
editor={Steven Harrison}
}

@book{coupland2014small,
title={Small talk},
author={Coupland, Justine},
year={2000},
pages={277},
publisher={Routledge},
address={New York}
}

@inproceedings{grevet2012eating,
title={Eating alone, together: new forms of commensality},
author={Grevet, Catherine and Tang, Anthony and Mynatt, Elizabeth},
booktitle={Proceedings of the 17th ACM international conference on Supporting group work},
pages={103--106},
year={2012}
}

@article{o2012food,
title={Food for talk: Phototalk in the context of sharing a meal},
author={O'Hara, Kenton and Helmes, John and Sellen, Abigail and Harper, Richard and Ten Bh{\"o}mer, Martijn and Van Den Hoven, Elise},
journal={Human-Computer Interaction},
volume={27},
number={1-2},
pages={124--150},
year={2012},
publisher={Taylor \& Francis}
}

@inproceedings{gallacher2015mood,
title={Mood squeezer: lightening up the workplace through playful and lightweight interactions},
author={Gallacher, Sarah and O'Connor, Jenny and Bird, Jon and Rogers, Yvonne and Capra, Licia and Harrison, Daniel and Marshall, Paul},
booktitle={Proceedings of the 18th ACM Conference on Computer Supported Cooperative Work \& Social Computing},
pages={891--902},
year={2015},
url = {https://doi-org.ucd.idm.oclc.org/10.1145/2675133.2675170},
doi = {10.1145/2675133.2675170}
}

@online{microsofttrends,
author = "Microsoft",
title = "{In Hybrid Work, Managers Keep Teams Connected}",
url={"https://www.microsoft.com/en-us/worklab/work-trend-index/managers-keep-teams-connected"},
year =  {2021},
month = {3}
}

@article{rockmann2015contagious,
title={Contagious offsite work and the lonely office: The unintended consequences of distributed work},
author={Rockmann, Kevin W and Pratt, Michael G},
journal={Academy of Management Discoveries},
volume={1},
number={2},
pages={150--164},
year={2015},
publisher={Academy of Management Briarcliff Manor, NY}
}

@inproceedings{cao2021large,
author = {Cao, Hancheng and Lee, Chia-Jung and Iqbal, Shamsi and Czerwinski, Mary and Wong, Priscilla N Y and Rintel, Sean and Hecht, Brent and Teevan, Jaime and Yang, Longqi},
title = {Large Scale Analysis of Multitasking Behavior During Remote Meetings},
year = {2021},
isbn = {9781450380966},
publisher = {Association for Computing Machinery},
address = {New York, NY, USA},
url = {https://doi-org.ucd.idm.oclc.org/10.1145/3411764.3445243},
doi = {10.1145/3411764.3445243},
booktitle = {Proceedings of the 2021 CHI Conference on Human Factors in Computing Systems},
articleno = {448},
numpages = {13},
location = {Yokohama, Japan},
series = {CHI '21}
}

@inproceedings{tang1994supporting,
title={Supporting distributed groups with a montage of lightweight interactions},
author={Tang, John C and Isaacs, Ellen A and Rua, Monica},
booktitle={Proceedings of the 1994 ACM conference on Computer supported cooperative work},
pages={13--34},
year={1994},
publisher={ACM New York, NY, USA},
url = {https://doi-org.ucd.idm.oclc.org/10.1145/192844.192861},
doi = {10.1145/192844.192861}
}

@misc{microsoft:pandemic,
title={A pulse on employees’ wellbeing, six months into the pandemic},
author={Jared Spataro},
howpublished = {blog},
year={2020},
month={9},
url={https://www.microsoft.com/en-us/microsoft-365/blog/2020/09/22/pulse-employees-wellbeing-six-months-pandemic/}
}

@inproceedings{affect_spot,
author = {Murali, Prasanth and Hernandez, Javier and McDuff, Daniel and Rowan, Kael and Suh, Jina and Czerwinski, Mary},
title = {AffectiveSpotlight: Facilitating the Communication of Affective Responses from Audience Members during Online Presentations},
year = {2021},
isbn = {9781450380966},
publisher = {Association for Computing Machinery},
address = {New York, NY, USA},
url = {https://doi-org.ucd.idm.oclc.org/10.1145/3411764.3445235},
doi = {10.1145/3411764.3445235},
booktitle = {Proceedings of the 2021 CHI Conference on Human Factors in Computing Systems},
articleno = {247},
numpages = {13},
location = {Yokohama, Japan},
series = {CHI '21}
}

@inproceedings{meeetingcoach,
author = {Samrose, Samiha and McDuff, Daniel and Sim, Robert and Suh, Jina and Rowan, Kael and Hernandez, Javier and Rintel, Sean and Moynihan, Kevin and Czerwinski, Mary},
title = {MeetingCoach: An Intelligent Dashboard for Supporting Effective \& Inclusive Meetings},
year = {2021},
isbn = {9781450380966},
publisher = {Association for Computing Machinery},
address = {New York, NY, USA},
booktitle = {Proceedings of the 2021 CHI Conference on Human Factors in Computing Systems},
articleno = {252},
numpages = {13},
location = {Yokohama, Japan},
series = {CHI '21},
url = {https://doi-org.ucd.idm.oclc.org/10.1145/3411764.3445615},
doi = {10.1145/3411764.3445615}
}

@inproceedings{follmer2010video,
title={Video play: playful interactions in video conferencing for long-distance families with young children},
author={Follmer, Sean and Raffle, Hayes and Go, Janet and Ballagas, Rafael and Ishii, Hiroshi},
booktitle={Proceedings of the 9th International Conference on Interaction Design and Children},
pages={49--58},
year={2010},
url = {https://doi-org.ucd.idm.oclc.org/10.1145/1810543.1810550},
doi = {10.1145/1810543.1810550}
}

@inproceedings{hunter2014waazam,
title={WaaZam! Supporting creative play at a distance in customized video environments},
author={Hunter, Seth E and Maes, Pattie and Tang, Anthony and Inkpen, Kori M and Hessey, Susan M},
booktitle={Proceedings of the 32nd annual ACM conference on Human factors in computing
systems},
pages={1197--1206},
year={2014}
}

@article{nonaka1998concept,
author = {Ikujiro Nonaka and Noboru Konno},
title ={The Concept of “Ba”: Building a Foundation for Knowledge Creation},
journal = {California Management Review},
volume = {40},
number = {3},
pages = {40-54},
year = {1998},
}

@inproceedings{karahalios2005chit,
title={Chit chat club: bridging virtual and physical space for social interaction},
author={Karahalios, Karrie G and Dobson, Kelly},
booktitle={CHI'05 extended abstracts on Human factors in computing systems},
pages={1957--1960},
year={2005},
url = {https://doi-org.ucd.idm.oclc.org/10.1145/1056808.1057066},
doi = {10.1145/1056808.1057066}
}

@inproceedings{foley2010netpot,
title={NetPot: easy meal enjoyment for distant diners},
author={Foley-Fisher, Zoltan and Tsao, Vincent and Wang, Johnty and Fels, Sid},
booktitle={International Conference on Entertainment Computing},
pages={446--448},
year={2010},
organization={Springer}
}

@book{salen2004rules,
title={Rules of play: Game design fundamentals},
author={Salen, Katie and Tekinba{\c{s}}, Katie Salen and Zimmerman, Eric},
year={2004},
publisher={MIT press}
}

@article{wiederhold2020connecting,
author = {Wiederhold, Brenda K.},
title = {Connecting Through Technology During the Coronavirus Disease 2019 Pandemic: Avoiding “Zoom Fatigue”},
journal = {Cyberpsychology, Behavior, and Social Networking},
volume = {23},
number = {7},
pages = {437-438},
year = {2020}
}

@inproceedings{anderson1999understanding,
title={Understanding multiparty multimedia interactions},
author={Anderson, AH and Mullin, J and Katsavaras, E and McEwan, R and Grattan, E and Brundell, P},
booktitle={Psychological Models of Communication in Collaborative Systems (AAAI Fall Symposium), North Falmouth, Massachusetts, United States},
address = {California: Menlo Park},
pages={9--16},
year={1999},
month={11}
}

@article{cao2021my,
title={My Team Will Go On: Differentiating High and Low Viability Teams through Team Interaction},
author={Cao, Hancheng and Yang, Vivian and Chen, Victor and Lee, Yu Jin and Stone, Lydia and Diarrassouba, N'godjigui Junior and Whiting, Mark E and Bernstein, Michael S},
journal={Proceedings of the ACM on Human-Computer Interaction},
volume={4},
number={CSCW3},
pages={1--27},
year={2021},
publisher={ACM New York, NY, USA}
}

@article{halford2005hybrid,
title={Hybrid workspace: Re-spatialisations of work, organisation and management},
author={Halford, Susan},
journal={New Technology, Work and Employment},
volume={20},
number={1},
pages={19--33},
year={2005},
publisher={Wiley Online Library}
}

@article{yang2020work,
author = {Yang, Longqi and Holtz, David and Jaffe, Sonia and Suri, Siddharth and Sinha, Shilpi and Weston, Jeffrey and Joyce, Connor and Shah, Neha Parikh and Sherman, Kevin and Hecht, Brent and Teevan, Jaime},
title = {The effects of remote work on collaboration among information workers},
year = {2021},
month = {9},
journal = {Nature Human Behaviour},
}

@article{kaushik2020impact,
title={The impact of pandemic COVID-19 in workplace},
author={Kaushik, Meenakshi and Guleria, Neha},
journal={European Journal of Business and Management},
volume={12},
number={15},
pages={9--18},
year={2020}
}

@inproceedings{stray2019slack,
author = {Stray, Viktoria and Moe, Nils Brede and Noroozi, Mehdi},
title = {Slack Me If You Can! Using Enterprise Social Networking Tools in Virtual Agile Teams},
year = {2019},
publisher = {IEEE Press},
booktitle = {Proceedings of the 14th International Conference on Global Software Engineering},
pages = {101–111},
numpages = {11},
 doi={10.1109/ICGSE.2019.00031}
}

@inproceedings{sarkar2021promise,
author = {Sarkar, Advait and Rintel, Sean and Borowiec, Damian and Bergmann, Rachel and Gillett, Sharon and Bragg, Danielle and Baym, Nancy and Sellen, Abigail},
title = {The Promise and Peril of Parallel Chat in Video Meetings for Work},
year = {2021},
isbn = {9781450380959},
publisher = {Association for Computing Machinery},
address = {New York, NY, USA},
booktitle = {Extended Abstracts of the 2021 CHI Conference on Human Factors in Computing Systems},
articleno = {260},
numpages = {8},
location = {Yokohama, Japan},
series = {CHI EA '21},
url = {https://doi-org.ucd.idm.oclc.org/10.1145/3411763.3451793},
doi = {10.1145/3411763.3451793}
}

@article{methot2020office,
author = {Methot, Jessica R. and Rosado-Solomon, Emily H and Downes, Patrick and Gabriel, Allison S},
title = {Office Chit-Chat as a Social Ritual: The Uplifting Yet Distracting Effects of Daily Small Talk at Work},
journal = {Academy of Management Journal},
year={2020},
volume = {0}
}

@inproceedings{rudnicka2020eworklife,
title={Eworklife: Developing effective strategies for remote working during the COVID-19 pandemic},
author={Rudnicka, Anna and Newbold, Joseph W and Cook, Dave and Cecchinato, Marta E and Gould, Sandy and Cox, AL},
year={2020},
booktitle={The New Future of Work Online Symposium},
publisher={Microsoft}
}

@article{doolittle2021association,
title={Association of burnout with emotional coping strategies, friendship, and institutional support among internal medicine physicians},
author={Doolittle, Benjamin R},
journal={Journal of clinical psychology in medical settings},
volume={28},
number={2},
pages={361--367},
year={2021},
publisher={Springer}
}

@article{kossek2016managing,
title={Managing work-life boundaries in the digital age},
author={Kossek, Ellen Ernst},
journal={Organizational Dynamics},
volume={45},
number={3},
pages={258--270},
year={2016},
publisher={Pergamon}
}

@inproceedings{kaur2020optimizing,
title={Optimizing for happiness and productivity: Modeling opportune moments for transitions and breaks at work},
author={Kaur, Harmanpreet and Williams, Alex C and McDuff, Daniel and Czerwinski, Mary and Teevan, Jaime and Iqbal, Shamsi T},
booktitle={Proceedings of the 2020 CHI Conference on Human Factors in Computing Systems},
pages={1--15},
year={2020},
url = {https://doi-org.ucd.idm.oclc.org/10.1145/3313831.3376817},
doi = {10.1145/3313831.3376817}
}

@inproceedings{koehne2012remote,
title={Remote and alone: coping with being the remote member on the team},
author={Koehne, Benjamin and Shih, Patrick C and Olson, Judith S},
booktitle={Proceedings of the ACM 2012 conference on Computer Supported Cooperative Work},
pages={1257--1266},
year={2012},
url = {https://doi-org.ucd.idm.oclc.org/10.1145/2145204.2145393},
doi = {10.1145/2145204.2145393}
}

@inproceedings{calefato2020case,
author = {Calefato, Fabio and Giove, Andrea and Lanubile, Filippo and Losavio, Marco},
title = {A Case Study on Tool Support for Collaboration in Agile Development},
year = {2020},
isbn = {9781450370936},
publisher = {Association for Computing Machinery},
address = {New York, NY, USA},
url = {https://doi-org.ucd.idm.oclc.org/10.1145/3372787.3390436},
doi = {10.1145/3372787.3390436},
booktitle = {Proceedings of the 15th International Conference on Global Software Engineering},
pages = {11–21},
numpages = {11},
location = {Seoul, Republic of Korea},
series = {ICGSE '20}
}

@article{de2020recover,
title={How to recover during and from a pandemic},
author={De Bloom, Jessica},
journal={Industrial Health},
volume={58},
number={3},
pages={197--199},
year={2020},
publisher={National Institute of Occupational Safety and Health}
}

@inproceedings{hayes2020m,
title={I'm not Working from Home, I'm Living at Work: Perceived Stress and Work-Related Burnout before and during COVID-19},
author={Hayes, Sherrill and Priestley, Jennifer L and Ishmakhametov, Namazbai and Ray, Herman E},
year={2020},
booktitle={PsyArXiv},
note={Unpublished Manuscript},
doi={10.31234/osf.io/vnkwa}
}

@book{guest2011applied,
  title={Applied thematic analysis},
  author={Guest, Greg and MacQueen, Kathleen M and Namey, Emily E},
  year={2011},
  publisher={SAGE Publications}
}

@article{gutwin2002descriptive,
  title={A descriptive framework of workspace awareness for real-time groupware},
  author={Gutwin, Carl and Greenberg, Saul},
  journal={Computer Supported Cooperative Work (CSCW)},
  volume={11},
  number={3},
  pages={411--446},
  year={2002},
  publisher={Springer}
}

@inproceedings{robert2018disaggregating,
author = {Robert, Lionel P. and You, Sangseok},
title = {Disaggregating the Impacts of Virtuality on Team Identification},
year = {2018},
isbn = {9781450355629},
publisher = {Association for Computing Machinery},
address = {New York, NY, USA},
url = {https://doi-org.ucd.idm.oclc.org/10.1145/3148330.3148337},
doi = {10.1145/3148330.3148337},
booktitle = {Proceedings of the 2018 ACM Conference on Supporting Groupwork},
pages = {309–321},
numpages = {13},
location = {Sanibel Island, Florida, USA},
series = {GROUP '18}
}

\newpage
\clearpage
\begin{figure*}[t]
    \centering
    \includegraphics[ width=\textwidth]{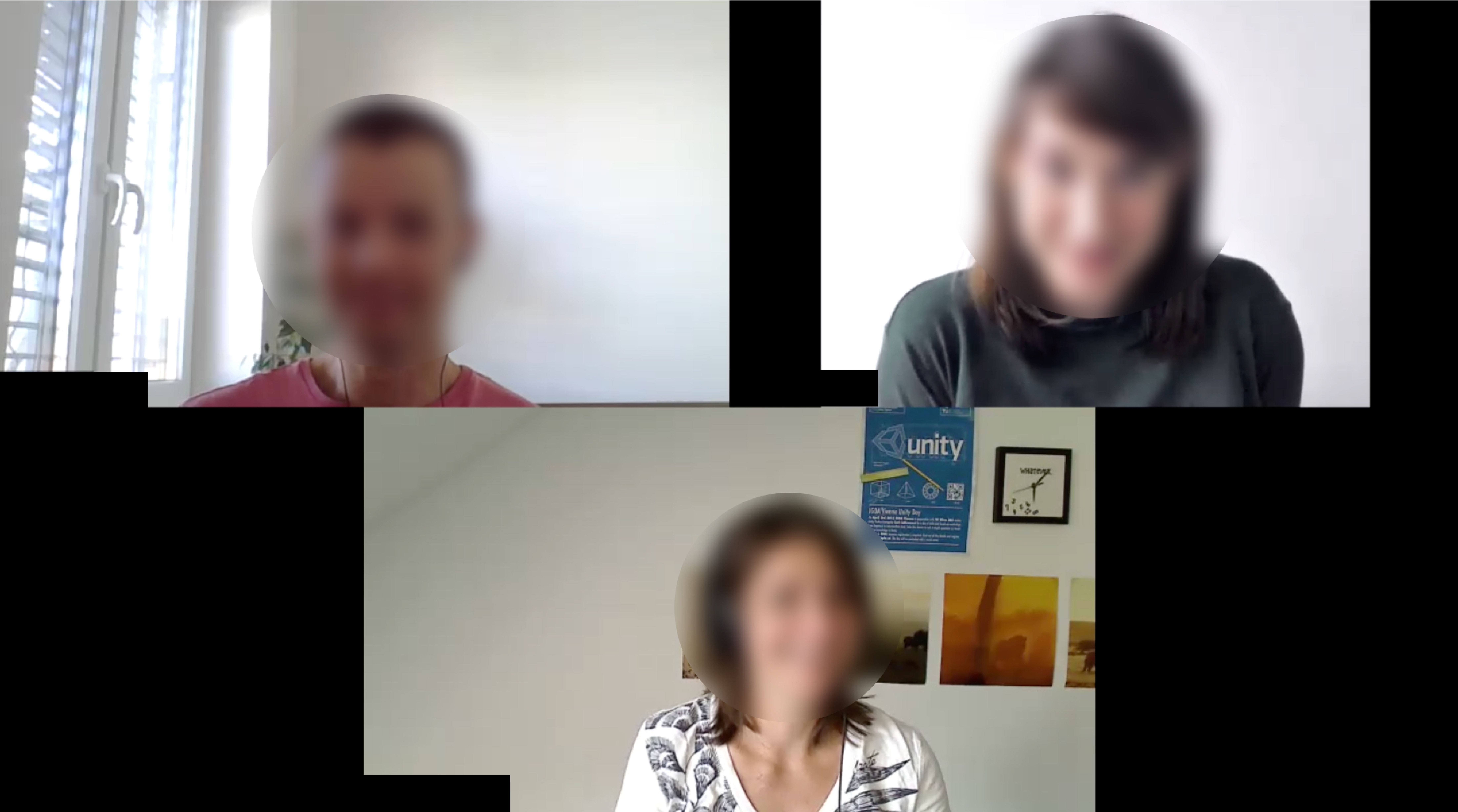}
    \caption{Example Participant Triad Conversing on Zoom}
    \label{fig:tasks}
\end{figure*}

\newpage
\clearpage

\begin{figure*}[t]
    \centering
    \includegraphics[width = \textwidth]{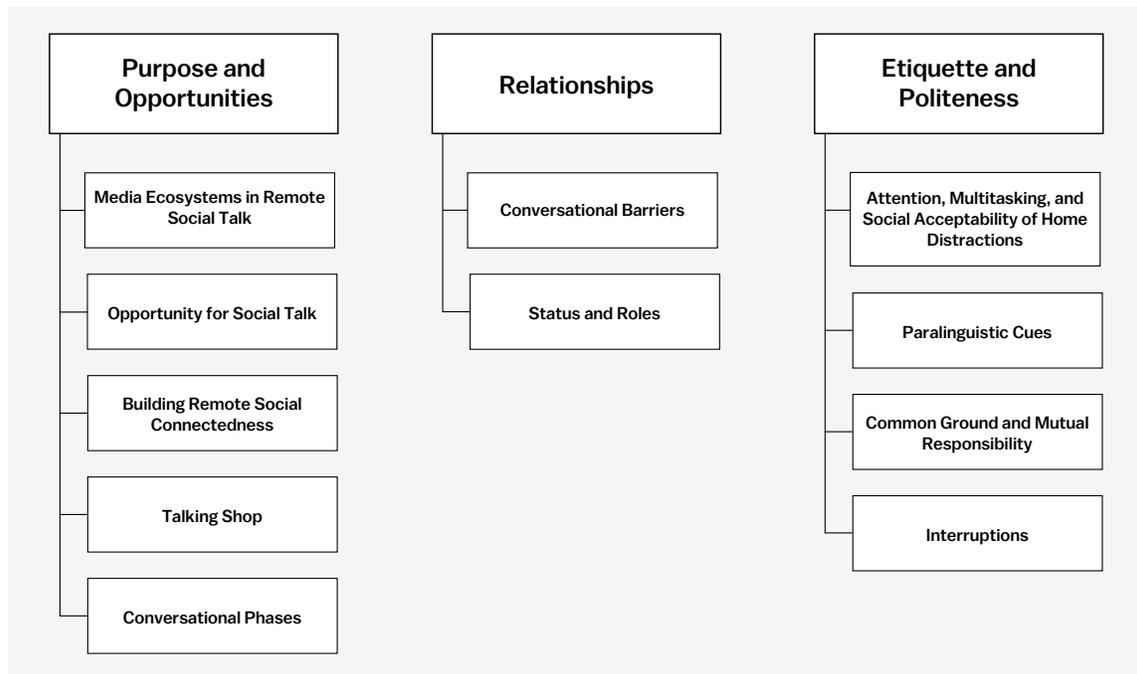}
    \caption{Themes and sub-themes from semi-structured interviews for social talk with colleagues via video conferencing during COVID-19}
    \label{fig:themes}
\end{figure*}

\end{document}